\documentclass[11pt,prd,onecolumn,showpacs,amsmath,amssymb,aps,floats,floatfix,nofootinbib]{revtex4-1}

\usepackage{hyperref} 

\usepackage{enumitem}
\usepackage{dcolumn}
\usepackage{bm}
\usepackage{amsmath}
\usepackage{amsfonts}
\usepackage{amssymb}
\usepackage{color}
\usepackage{latexsym}
\usepackage{slashed} 
\usepackage{pstricks}
\usepackage{indentfirst}
\usepackage{mathrsfs}
\usepackage{multirow}
\usepackage{epsfig,psfrag}
\usepackage{subfigure}
\usepackage{setspace} 
\usepackage[scientific-notation=true]{siunitx} 

\graphicspath{{figs/}}

\def\ifb{\mathrm{fb}^{-1}} 
\def\GeV{\mathrm{GeV}}     
\def\tchi{{\tilde \chi}}  
\def\smur{{\tilde {\mu}}_\mathrm{R}}  
\def\pT{p_\mathrm{T}} 
\def\missET{\slashed E_\mathrm{T}} 
\def\missp{\slashed{p}} 

\def\chia{\tilde{\chi}_1^0}

\def\tchi{{\tilde\chi}}  
\def\mY{m_Y^\mathrm{edge}}
\def\mN{m_N^\mathrm{edge}}

\renewcommand{\vec}[1]{\mathbf{#1}}

\makeatother

\allowdisplaybreaks 

\begin{document}
\allowdisplaybreaks 
\setlength\arraycolsep{0.2em} 
\setstretch{1.2} 

\title{Measuring Masses in Semi-Invisible Final States at Electron-Positron Colliders}
\author{Qian-Fei Xiang$^{1,2}$}
\author{Xiao-Jun Bi$^1$}
\author{Qi-Shu Yan$^{2,3}$}
\author{Peng-Fei Yin$^1$}
\author{Zhao-Huan Yu$^4$}
\affiliation{$^1$Key Laboratory of Particle Astrophysics,
Institute of High Energy Physics, Chinese Academy of Sciences,
Beijing 100049, China}
\affiliation{$^2$School of Physical Sciences,
University of Chinese Academy of Sciences,
Beijing 100049, China}
\affiliation{$^3$Center for Future High Energy Physics, Chinese Academy of Sciences, Beijing 100049, China}
\affiliation{$^4$ARC Centre of Excellence for Particle Physics at the Terascale,
School of Physics, The~University of Melbourne, Victoria 3010, Australia}

\begin{abstract}
Mass measurement of a particle whose decay products including invisible particles is a challenging task at colliders.
For a new physics model involving a dark matter candidate $N$ and a $Z_2$ symmetry that stabilizes it, a typical new process at $e^+e^-$ colliders is the pair production $e^+e^- \to Y\bar{Y}$ followed by decay processes $Y\to aN$ and $\bar{Y}\to b\bar{N}$, where $a$ and $b$ are visible but $N$ is invisible.
In this work, we propose a new method to measure the physical masses in this topology by making use of the kinematic equations given by momentum-energy conservation and on-shell conditions.
For each event, the solvability of these equations determines a limited region on the trial $m_Y$-$m_N$ plane.
The edge of this region can be used to define two variables, $\mY$ and $\mN$, whose distributions are utilized to derive the measurement values of $m_Y$ and $m_N$.
The measurement deviations and uncertainties are also estimated after including detector effects and background contamination.
\end{abstract}


\maketitle

\section{Introduction}
The discovery of the Higgs boson at the LHC~\cite{Aad:2012tfa,Chatrchyan:2012xdj} is a great triumph of the Standard Model (SM).
Nevertheless, plenty of astrophysical observations indicate that the SM is incomplete, because it is unable to provide a suitable candidate for cold dark matter (DM).
Among various possibilities, DM models containing weakly interacting massive particles, such as supersymmetric models and universal extra dimensional models, are likely to interpret the observational results better and are, hence, more attractive.
In the view of model building, DM particles are often associated by extra new particles at the TeV scale, which typically decay into visible SM particles and DM particles that are invisible in a general-purpose detector.
Therefore, at high energy colliders, their production processes would induce missing energy signatures, which are the primary hint of DM production at the LHC and future colliders.

In order to reveal the nature of DM, it will be essential to measure the properties of the new particles, such as mass, spin, parity, and other quantum numbers.
In this work, we focus on the mass measurement of invisible particles as well as their parents at high energy colliders.
Along this line of thinking, many efficient techniques have been proposed and developed in the literature (see Refs.~\cite{Barr:2010zj,Kong:2013xma} for reviews), such as endpoint methods~\cite{Hinchliffe:1996iu,Bachacou:1999zb,Hinchliffe:1999zc,Allanach:2000kt,Gjelsten:2004ki,Gjelsten:2005aw,Miller:2005zp,Burns:2008cp}, $M_\mathrm{T2}$ methods~\cite{Lester:1999tx,Barr:2003rg,Meade:2006dw,Cho:2007qv,Barr:2007hy,Lester:2007fq,Gripaios:2007is,Nojiri:2008vq,Tovey:2008ui,Burns:2008va,Barr:2011xt},
polynomial methods~\cite{Nojiri:2003tu,Kawagoe:2004rz,Cheng:2008mg,Cheng:2009fw,Cheng:2007xv}, and hybrid methods~\cite{Nojiri:2007pq,Ross:2007rm,Barr:2008ba,Cho:2008tj,Cheng:2008hk}. The basic idea of many methods is to resolve the mass of an invisible particle by using the momentum-energy conservation and some kinematic conditions, which require that the related particles should be close to mass shells. For instance, endpoint methods attempt to determine the endpoints of invariant mass distributions. By combining several endpoints, the mass of the invisible particle could be determined. $M_\mathrm{T2}$ methods can give a lower bound on the mass of a parent particle in the decay chain, and is widely used to determine the masses of known particles as well as to explore new physics.
Recently, some new methods for measuring masses of invisible particles have also been proposed~\cite{Han:2012nm,Han:2012nr,Agashe:2012fs,Agashe:2013eba,Konar:2008ei,Konar:2010ma,Cho:2015laa,Lim:2016ymd,Edelhauser:2012xb,Mahbubani:2012kx,Cho:2014naa,Cho:2014yma,Swain:2014dha,Konar:2015hea,Konar:2016wbh,Agashe:2012bn,Agashe:2015wwa,Agashe:2015ike}.
For instance, inspired by the endpoint methods, some methods focus on other local kinematic features in the distributions, such as cusps~\cite{Han:2012nm,Han:2012nr} and peaks~\cite{Agashe:2012fs,Agashe:2013eba}.

At hadron colliders, the longitudinal momentum of the initial state is unknown and only the transverse components of the missing momentum can be reconstructed through the measurement of visible particles in the final state based on the transverse momentum conservation. Therefore, only the transverse mass of an invisible particle can be extracted, and it always suffers a large uncertainty from the pollution of messy background processes, such as underlying events and pileup effects. In contrast, at lepton colliders, the well-measured energies and momenta of both initial and final states can directly determine the missing energy and the missing longitudinal momentum. In other words, there are two more kinematic equations that can be used to derive the unknown masses. Many mass measurement techniques at lepton colliders have been proposed~\cite{Conley:2010jk,Christensen:2014yya,HarlandLang:2012gn}. These methods utilize the kinematic features of the event distributions, such as endpoints and cusps~\cite{Conley:2010jk,Christensen:2014yya}, or focus on directly solving the kinematic conservation equations ~\cite{HarlandLang:2012gn}.

Generally speaking, the kinematics of new particles are determined by the unknown particle spectra and interactions in a new physics model, and thus some mass measurement methods may be quite specific. Nevertheless, since the kinematic conditions may not significantly depend on the details of the underlying models, many methods can be generalized. In a typical DM model, invisible DM particles can be produced in cascade decays of heavier particles. It is interesting to observe that the topology of the decay chains would determine the kinematic features of the final state.

In this work, we consider the simplest topology where a pair of heavy particles (denoted by $Y$) are produced via the process $e^+ e^- \to Y \bar{Y}$ and each of them further decays into a visible SM particle and an invisible particle (denoted by $N$). Then the process becomes $e^+e^- \to Y \bar{Y} \to a b N\bar{N}$, where $a$ and $b$ are visible, but $N$ and $\bar{N}$ are invisible.
We will propose a new mass measurement method for this topology at future high energy electron-positron colliders, e.g. the Circular Electron Positron Collider~\cite{CEPC}, International Linear Collider~\cite{Fujii:2015jha}, and Future Circular Collider~\cite{d'Enterria:2016cpw}. It can be realized in many typical new physics models, for instance, $e^+ e^- \to {\tilde\ell}^+{\tilde\ell}^- \to \ell^+ \ell^- \chia \chia$ in supersymmetric models and $e^+ e^- \to \ell_1^+ \ell_1^- \to \ell^+ \ell^- \gamma_1 \gamma_1$ in the universal extra dimensional model. The method is simply based on the solvability of kinematic equations.

There are eight unknown variables to describe the 4-momenta of two invisible particles, but only six kinematic constraint equations are available. However, if two trial masses of $Y$ and $N$ are introduced, these equations may be solved.
Therefore, for each event, there is a solvable region on the $m_Y^\mathrm{trial}$-$m_N^\mathrm{trial}$ plane, from which some new variables
can be constructed to extract the true masses $m_Y^\mathrm{true}$ and $m_N^\mathrm{true}$. This strategy has been proposed in Ref.~\cite{HarlandLang:2012gn}. In that work, the authors defined two variables $m_Y^\mathrm{max}$ and $m_N^\mathrm{max}$, which are the maximum values of allowed $m_Y^\mathrm{trial}$ and $m_N^\mathrm{trial}$ in the solvable region, respectively. These variables can be analytically calculated; the true masses $m_Y^\mathrm{true}$ and $m_N^\mathrm{true}$ can be efficiently obtained from the sharp edges of the $m_Y^\mathrm{max}$ and $m_N^\mathrm{max}$ distributions.

In this work, we propose two new variables $\mY$ and $\mN$, which are the coordinates of the furthest point in the solvable region from the origin point. The point of ($\mY$, $\mN$) is different from that of ($m_Y^\mathrm{max}$, $m_N^\mathrm{max}$) on the trial mass plane. If the true masses $m_Y^\mathrm{true}$ and $m_N^\mathrm{true}$ are large enough, the sharp structures of the 1D differential distributions of these variables can be used to obtain the true masses. However, if $m_Y^\mathrm{true}$ or $m_N^\mathrm{true}$ are small, the edges may not be sharp.
In our study, we find that the peak structure in the 2D density distribution of ($\mY$, $\mN$) would be very useful with sufficient events. We then use this feature to extract the true masses. Moreover, we estimate the deviation and statistical uncertainty for the mass measurement at $e^+e^-$ colliders with $\sqrt{s} = 500$ and $240$~GeV, including simplified detector effects and background contamination.

This paper is organized as follows.
In Sec.~\ref{sec:eq}, we briefly describe the kinematic equations due to momentum-energy conservation and on-shell conditions.
Based on the solvability of kinematic equations, we introduce two new variables ($\mY$, $\mN$) in Sec.~\ref{sec:new}.
In Sec.~\ref{sec:realistic} we study the mass measurement by utilizing the distributions of $\mY$ and $\mN$ with realistic considerations at future $e^+e^-$ colliders.
We end this paper with conclusions and discussions in Sec.~\ref{sec:cad}.

\section{Kinematic equations}
\label{sec:eq}

\begin{figure}[!htbp]
\centering
\includegraphics[width=.5\textwidth]{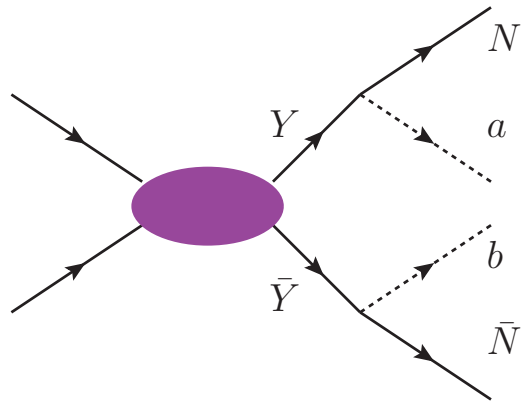}
\caption{Topology of the pair production of $Y$ and $\bar{Y}$, which subsequently decay into two visible particles ($a$ and $b$) and two invisible particles ($N$ and $\bar{N}$).}
\label{fig:prod}
\end{figure}

In this work, we consider a pair production process of heavy particles $Y$ and $\bar{Y}$ at $e^+e^-$ colliders. As shown in Fig.~\ref{fig:prod}, each of them subsequently decays into a visible particle ($a$ or $b$) and an invisible particle ($N$ or $\bar{N}$):
\begin{equation}
Y \to a({p_a}) + N({k_1}),\quad \bar Y \to b({p_b}) + \bar N({k_2}),
\end{equation}
where  $p_a$, $p_b$, $k_1$, and $k_2$ denote the 4-momenta of $a$, $b$, $N$, and $\bar{N}$, respectively.
The kinematic equations due to momentum-energy conservation and on-shell conditions are given by
\begin{eqnarray}
&& {q^\mu } = p_a^\mu  + p_b^\mu  + k_1^\mu  + k_2^\mu, \quad \mu = 0,1,2,3,
\label{eq:4mom}\\
&& k_1^2 = k_2^2 = m_N^2,
\label{eq:N_OS}\\
&& {({p_a} + {k_1})^2} = {({p_b} + {k_2})^2} = m_Y^2,
\label{eq:Y_OS}
\end{eqnarray}
where ${q^\mu } = (\sqrt{s},0,0,0)$ is the 4-momenta of the collision system.

At hadron colliders, $q^0$ and $q^3$ are unknown, and thus two equations in Eq.~\eqref{eq:4mom} are unavailable. Therefore, the components of $k_1^\mu$ and $k_2^\mu$, which are eight variables, cannot be directly obtained by just six kinematic equations even with $m_Y$ and $m_N$ as input parameters. In contrast, at $e^+e^-$ colliders, as the full 4-momentum of the collision system can be well determined, $k_1^\mu$ and $k_2^\mu$ may be unambiguously determined by solving the eight kinematic equations with trial values of $m_Y$ and $m_N$ given.
It is interesting to note that given the measured 4-momenta $p_a^\mu$, $p_b^\mu$ and $q^\mu$ in each event, the kinematic equations can be solved not only for the true physical masses $(m_Y^\mathrm{true}, m_N^\mathrm{true})$, but also for many sets of $(m_Y, m_N)$; in other words, there is a solvable region on the $(m_Y, m_N)$ plane for a given kinematics.

Below we illustrate how to extract $(m_Y^\mathrm{true}, m_N^\mathrm{true})$ from the region allowed by the kinematic equations. For this purpose,
we employ \texttt{MadGraph~5}~\cite{Alwall:2014hca} to generate simulation samples for the SM process $e^+e^-\to W^+W^-\to \mu^+\mu^-\nu\bar\nu$ and the supersymmetric process $e^+e^-\to \smur^+\smur^- \to \mu^+\mu^-\tchi_1^0\tchi_1^0$ with $m_{\tchi_1^0}=115~\GeV$ and $m_{\smur}=175~\GeV$ at $\sqrt{s}=500~\GeV$. For each event, $p_a^\mu$ and $p_b^\mu$ are given by the 4-momenta of the muons; we solve the kinematic equations and attempt to obtain the values of $k_1^\mu$ and $k_2^\mu$ for trial values of $(m_Y, m_N)$.

\begin{figure}[!htbp]
\centering
\subfigure[~$e^+e^-\to W^+W^-\to \mu^+\mu^-\nu\bar\nu$
\label{fig:sol_prob:ww}]
{\includegraphics[width=.45\textwidth]{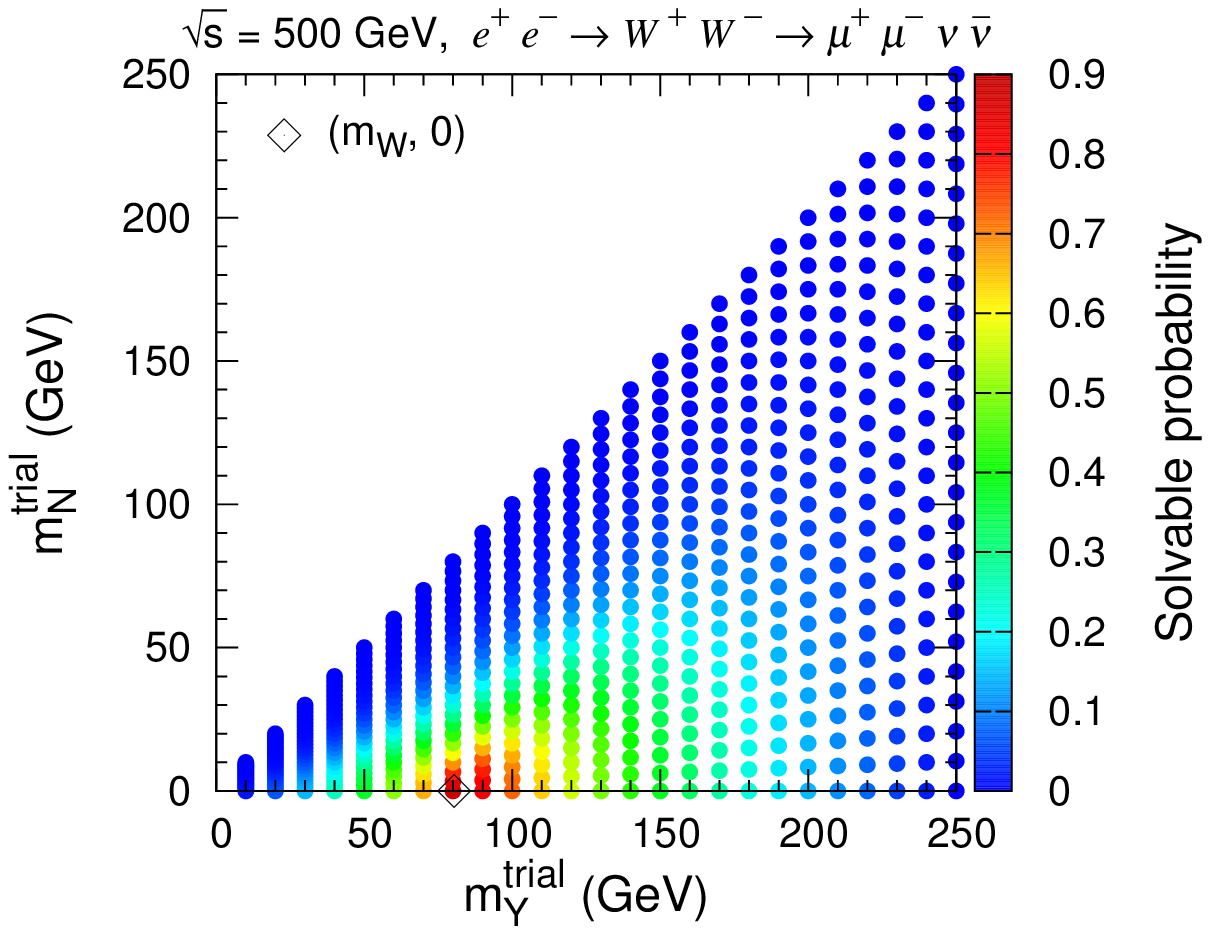}}
\hspace*{0.01\textwidth}
\subfigure[~$e^+e^-\to \smur^+\smur^- \to \mu^+\mu^-\tchi_1^0\tchi_1^0$
\label{fig:sol_prob:smur}]
{\includegraphics[width=.45\textwidth]{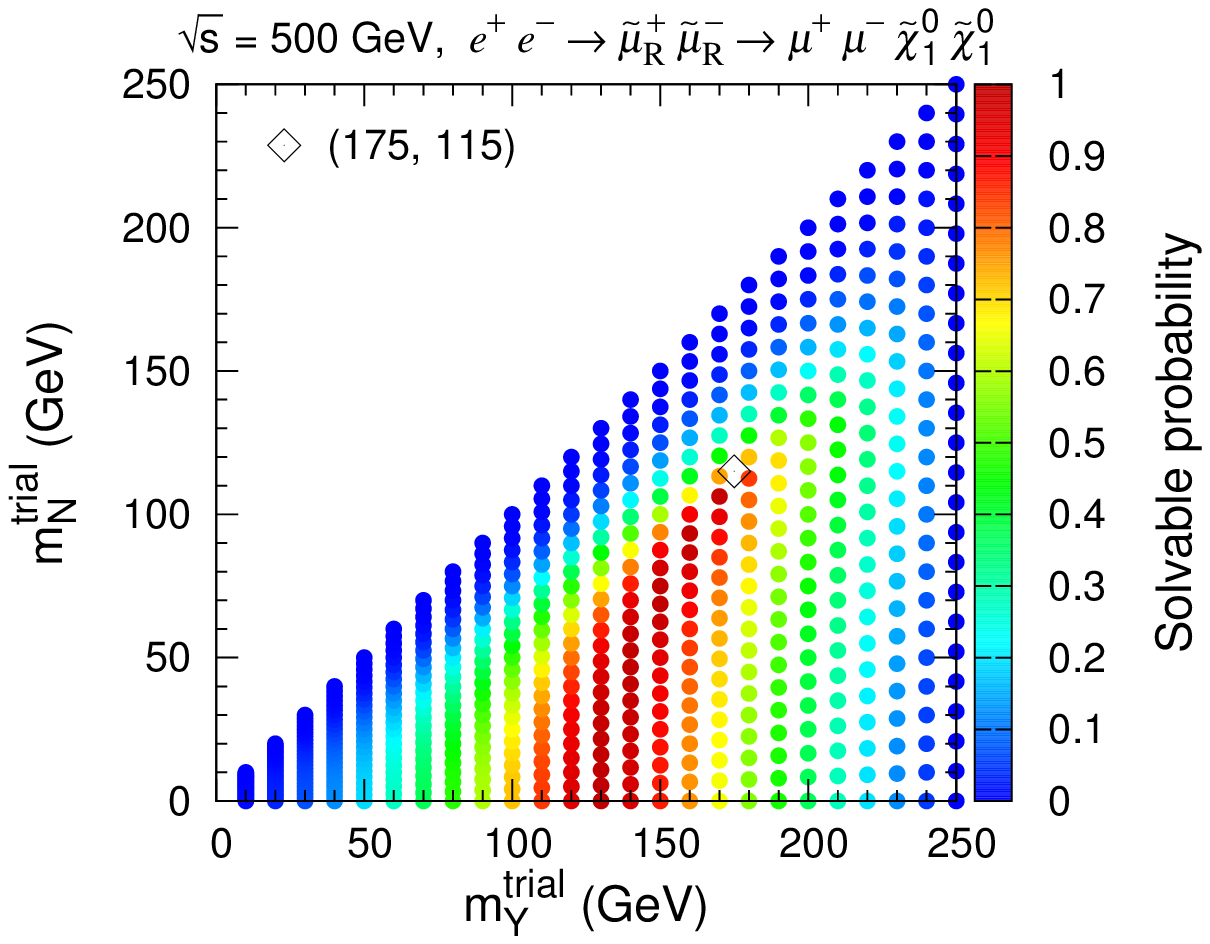}}
\caption{Solvable probability of trial values of $(m_Y, m_N)$ for $e^+e^-\to W^+W^-\to \mu^+\mu^-\nu\bar\nu$ (a) and $e^+e^-\to \smur^+\smur^- \to \mu^+\mu^-\tchi_1^0\tchi_1^0$ (b).
The empty diamonds denote the true values of $(m_Y, m_N)$.}
\label{fig:sol_prob}
\end{figure}

Apparently, given the true values of $(m_Y, m_N)$, the kinematic equations must have solutions for all events. Meanwhile, it should be noticed that for some events other trial values can also solve the equations. Therefore, for each point on the $m_Y^\mathrm{trial}$-$m_N^\mathrm{trial}$ plane, we can estimate a fraction of the events for which the kinematic equations are solvable and call it the ``solvable probability'', whose distributions are demonstrated in Fig.~\ref{fig:sol_prob}. From Fig.~\ref{fig:sol_prob:smur} for $e^+e^-\to \smur^+\smur^- \to \mu^+\mu^-\tchi_1^0\tchi_1^0$, we find that the point corresponding to the true masses is located at the edge of the region with a large solvable probability. But this is not obvious in Fig.~\ref{fig:sol_prob:ww} for $e^+e^-\to W^+W^-\to \mu^+\mu^-\nu\bar\nu$, because $m_N = 0$ squeezes the solvable region to be very small.

As we have mentioned, for each event, there is a region on the $m_Y$-$m_N$ plane inside which the kinematic equations have solutions. Moreover, the above analysis about the solvable probability tells us that the true masses tend to lie on the edge of this region and to be far away from the origin $(0,0)$.
Based on this observation, for each event we denote the point in the solvable region that has the largest distance from the origin as $(\mY, \mN)$.
Here the distance from the origin is defined as $\sqrt{m_Y^2+m_N^2}$.
Then it is expected that $\sqrt{(m_Y^{\rm{edge}})^2+(m_N^{\rm{edge}})^2}\geq \sqrt{(m_Y^{\rm{true}})^2+(m_N^{\rm{true}})^2}$.

\section{New kinematic variables}
\label{sec:new}
In this section, we give the realistic definition of $(\mY, \mN)$. The solvable region can be completely determined by the measured 4-momenta $p_a^\mu$  and $p_b^\mu$. Here we define several dimensionless variables normalized by $\sqrt{s}$ in the center-of-mass frame: $z_a \equiv p_a^0/\sqrt{s}$, $z_b \equiv p_b^0/\sqrt{s}$, $\vec{a} \equiv \vec{p}_a/\sqrt{s}$, $\vec{b} \equiv  \vec{p}_b/\sqrt{s}$, $\mu_N \equiv m_N/\sqrt{s}$, and $\mu_Y \equiv m_Y/\sqrt{s}$.
The normalized 3-momentum and energy of one invisible particle $N$ are defined as  $\vec{k} \equiv \vec{k}_{1}/\sqrt{s}$ and $z \equiv k_1^0/\sqrt{s}$.
By using the momentum-energy conservation equation~\eqref{eq:4mom}, the normalized 3-momentum and energy of the other invisible particle are given by $\vec{k}' \equiv \vec{k}_{2}/\sqrt{s} = -\vec{k}-\vec{a}-\vec{b}$ and $z' \equiv k_2^0/\sqrt{s} = 1-z_a -z_b-z$.
Thus, the four on-shell conditions in Eqs.~\eqref{eq:N_OS} and \eqref{eq:Y_OS} can be expressed as
\begin{eqnarray}
	|\vec{k}|^2 + \mu_N^2 &= &z^2 = (z_Y -z_a)^2, \label{eq:on1}\\
	|\vec{k} +\vec{a} +\vec{b}|^2 + \mu_N^2 &=& (1-z_a-z_b-z)^2, \label{eq:on2}\\
	|\vec{k}+\vec{a}|^2 + \mu_Y^2 &=& z_Y^2,\label{eq:on3}\\
	|\vec{k}+\vec{a}|^2 +\mu_Y^2 &=& (1-z_Y)^2, \label{eq:on4}
\end{eqnarray}
where $z_Y$ is the normalized energy of the intermediate particle $Y$.

The on-shell condition~\eqref{eq:on1} can be rearranged as
\begin{equation}
	\vec{k} \cdot \vec{k} = K \label{eq:dis},
\end{equation}
with
\begin{equation}
	K \equiv \left(\frac{1}{2}-z_a\right)^2 -\mu_N^2.
	\label{eq:A}
\end{equation}
Eliminating $|\vec{k}|^2$ in Eqs.~\eqref{eq:on2}--\eqref{eq:on4}, we obtain two equations depending on $\vec{k}$:
\begin{eqnarray}
	\label{eq:sol}
	\vec{a} \cdot \vec{k} = A, \quad
	\vec{b} \cdot \vec{k} = B,
\end{eqnarray}
where
\begin{equation}
A \equiv \frac{1}{2} (z_a -z_a^2 -\mu_Y^2 +\mu_N^2 - |\vec{a}|^2),\quad
B \equiv \frac{1}{2} (z_b^2-z_b +\mu_Y^2 - \mu_N^2 - |\vec{b}|^2) - \vec{a} \cdot \vec{b}.
\label{eq:BC}
\end{equation}
By using Eq.~\eqref{eq:sol}, the second and third components of $\vec{k}$, $k_y$ and $k_z$, can be expressed by the first component $k_x$. Then Eq.~\eqref{eq:dis} leads to an quadratic equation of $k_x$:
\begin{eqnarray}
&& |\vec{a}\times \vec{b}|^2 k_x^2 +2 \left[ (A b_z -B a_z)(a_z b_x-a_x b_z) + (A b_y - B a_y)(a_y b_x-a_x b_y) \right] k_x \nonumber\\
&& \quad + (A b_z - B a_z)^2 + (A b_y - B a_y)^2 - K (a_z b_y - a_y b_z)^2 =0.
\label{eq:kx}
\end{eqnarray}
The solvable condition for this equation can be written in a compact form as
\begin{equation}
	\sqrt{K} \geq \frac{|A \vec{b} -B \vec{a}|}{|\vec{a} \times \vec{b}|}.
	\label{eq:judge}
\end{equation}

The inequality~\eqref{eq:judge} has a geometrical explanation.
Two equations in Eq.~\eqref{eq:sol} represent two planes in the three-dimensional $\vec{k}$ space, which are perpendicular to $\vec{a}$ and $\vec{b}$, respectively.
The values of $\vec{k}$ allowed by Eq.~\eqref{eq:sol} should be located on the line of intersection between two planes. The right-hand side of the inequality~\eqref{eq:judge} is the distance from the origin to this line, while $\sqrt{K}$ is the radius of the sphere that is described by Eq.~\eqref{eq:dis} and is centered at the origin.
Therefore, the inequality~\eqref{eq:judge} just means that the line defined by Eq.~\eqref{eq:sol} should intersect the sphere to give real solutions.

Substituting Eqs.~\eqref{eq:A} and \eqref{eq:BC} into the inequality~\eqref{eq:judge}, we derive an inequality for $\mu_Y^2$ and $\mu_N^2$:
\begin{equation}
	A_0 (\mu_Y^2 - \mu_N^2)^2 + B_0 (\mu_Y^2 - \mu_N^2) + C_0 \mu_N^2 + D_0 \leq 0,
	\label{eq:ineq}
\end{equation}
where the coefficients $A_0$, $B_0$, $C_0$, and $D_0$ are given by
\begin{eqnarray}
	A_0 &\equiv&|\vec{a} + \vec{b}|^2,\\
	B_0 &\equiv& 2 |\vec{a}|^2 (z_b^2 - z_b) + 2 |\vec{b}|^2 (z_a^2-z_a) +2 \vec{a}\cdot\vec{b} (z_a^2 + z_b^2 -z_a -z_b - |\vec{a} +\vec{b}| ^2),\\
	C_0 &\equiv& 4 |\vec{a} \times \vec{b}|^2,\\ \nonumber
	D_0&\equiv& |\vec{a}|^2 |\vec{b}|^2 |\vec{a}+\vec{b}|^2 - |\vec{a} \times \vec{b}|^2 + 2 |\vec{a}|^2 |\vec{b}|^2 (z_a + z_b -z_a^2 -z_b^2) \\
	&& +|(z_b^2-z_b)\vec{a}+(z_a^2-z_a) \vec{b}|^2 -2 \vec{a}\cdot \vec{b} \left[  |\vec{a}|^2(z_b^2-z_b)+|\vec{b}|^2 (z_a^2-z_a) \right].
\end{eqnarray}
Note that the 4-vectors $(z_a,\vec{a})$ and $(z_b,\vec{b})$ are exchangeable in these coefficients.

\begin{figure}[!htbp]
\centering
\includegraphics[width=.45\textwidth]{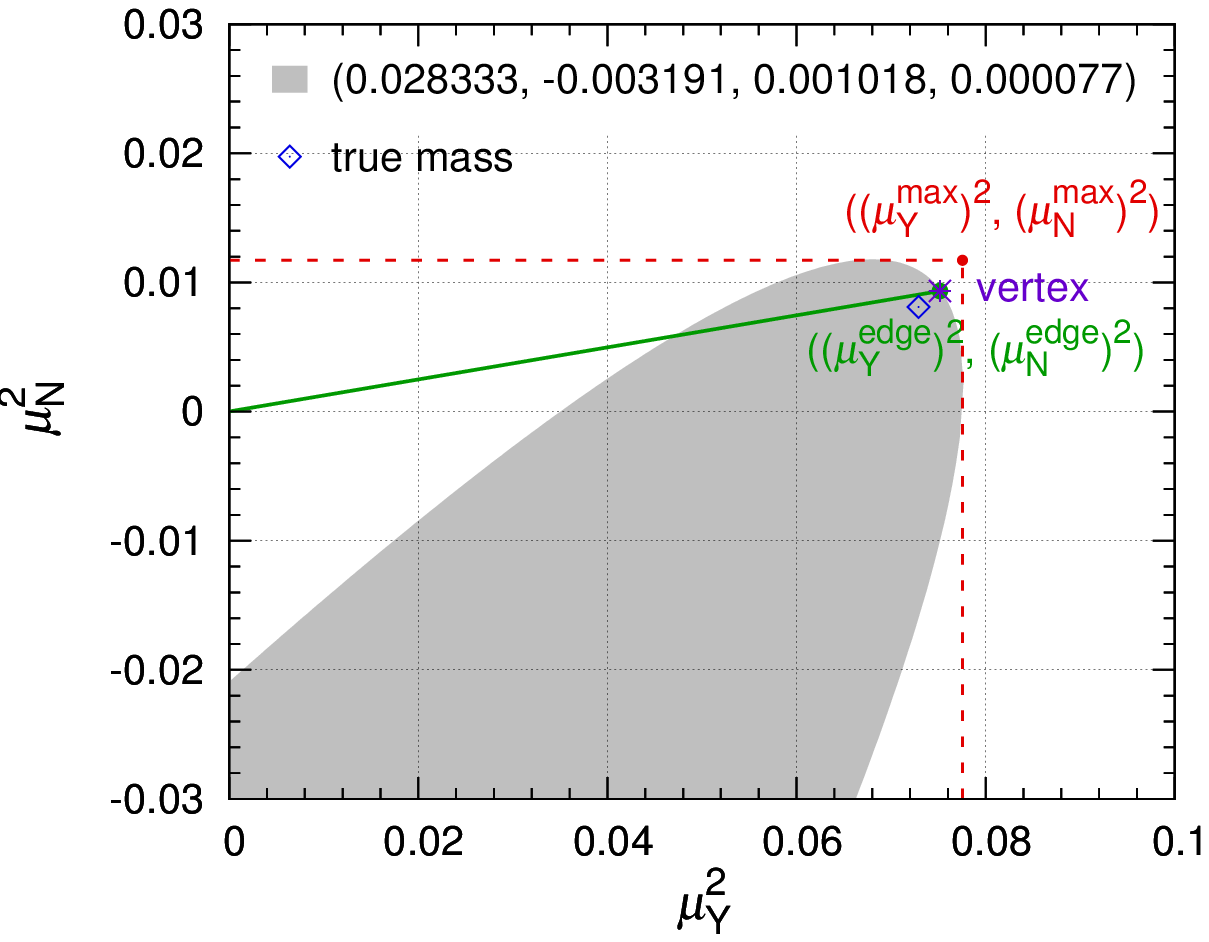}
\hspace{.5em}
\includegraphics[width=.45\textwidth]{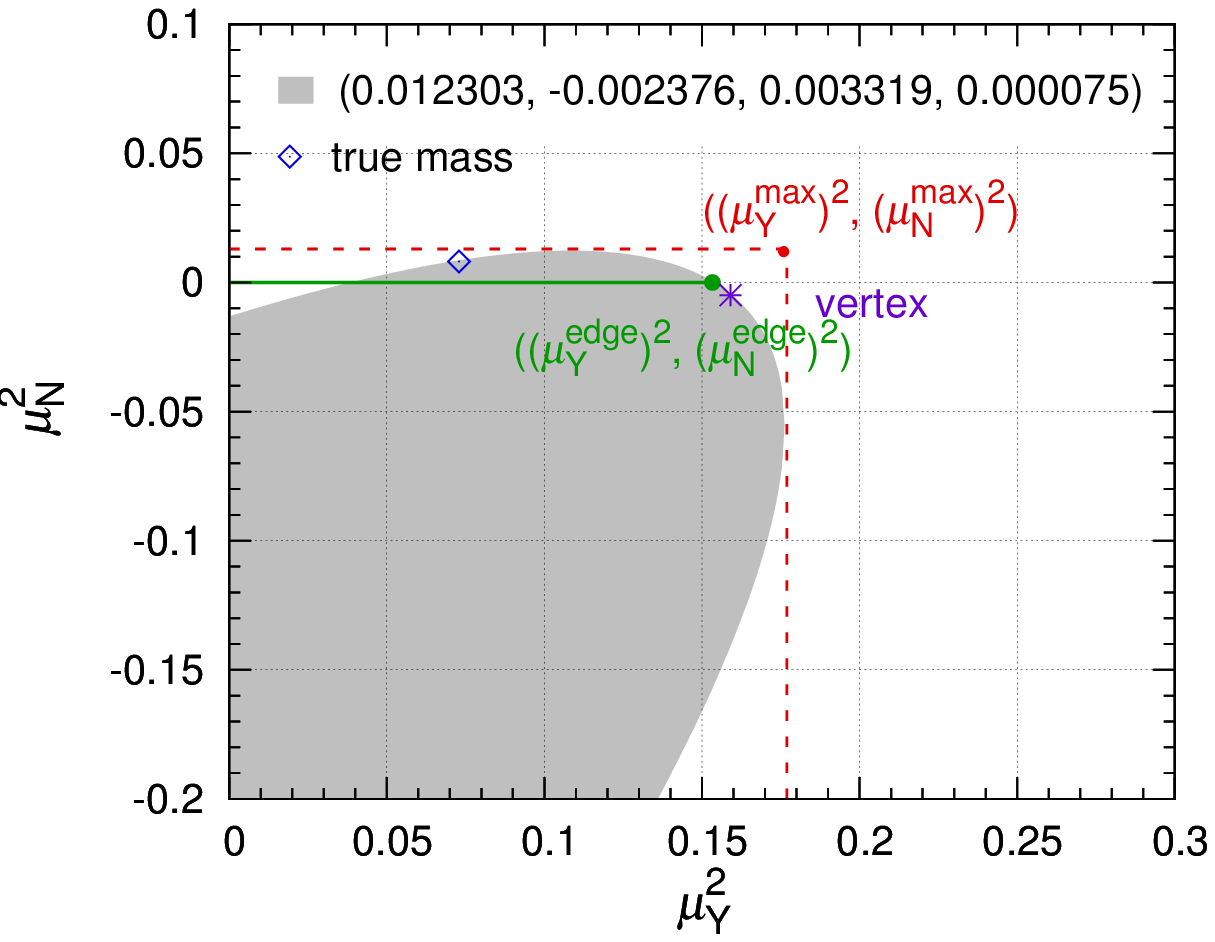}
\caption{Illustration of the solvable regions on the $\mu_Y^2-\mu_N^2$ plane derived from the inequality~\eqref{eq:ineq} for two specific events. The values of $A_0$, $B_0$, $C_0$, and $D_0$ are given in the legends. The gray color denotes the solvable regions, which are bounded by parabolas.
In the left panel, the vertex of the parabola is located in quadrant I and coincides with the point $\big((\mu_Y^\mathrm{edge})^2, (\mu_N^\mathrm{edge})^2\big)$.
In the right panel, the vertex is located in Quadrant IV and the point $\big((\mu_Y^\mathrm{edge})^2, (\mu_N^\mathrm{edge})^2\big)$ is located on the line $\mu_N^2=0$. The true mass point and the point $\big((\mu_Y^\mathrm{max})^2, (\mu_N^\mathrm{max})^2\big)$ are also denoted.}
\label{fig:sketch}
\end{figure}

The inequality~\eqref{eq:ineq} means that the solvable region for each event is bounded by a parabola on the $(\mu_Y^2, \mu_N^2)$ plane. The axis of symmetry of this parabola has a slope of $1$, so it is parallel to the line $\mu_Y^2=\mu_N^2$. Fig.~\ref{fig:sketch} shows the solvable regions on the $\mu_Y^2-\mu_N^2$ plane for two specific events. If the vertex of the parabola is located in quadrant I, as in the left panel of Fig.~\ref{fig:sketch}, it will be the furthest physical point from the origin in the solvable region. Therefore, by defining $\mu_Y^\mathrm{edge}\equiv m_Y^\mathrm{edge}/\sqrt{s}$ and $\mu_N^\mathrm{edge}\equiv m_N^\mathrm{edge}/\sqrt{s}$, we identify the vertex as the point $\big((\mu_Y^\mathrm{edge})^2, (\mu_N^\mathrm{edge})^2\big)$, whose values are given by
\begin{equation}
	(\mu_Y^\mathrm{edge})^2 =  \frac{4 B_0^2+3 C_0^2 -16 A_0 D_0 - 8 B_0 C_0}{16 A_0 C_0},\quad
	(\mu_N^\mathrm{edge})^2 = \frac{4 B_0^2 - C_0^2 - 16 A_0 D_0}{16 A_0 C_0}.
\end{equation}
If the vertex is located in quadrant IV, as in the right panel of Fig.~\ref{fig:sketch}, the furthest physical point will be the intersecting point of the parabola and the $\mu_Y^2$ axis, because the physical $\mu_Y^2$ should not be negative. In this case, we have
\begin{equation}
	(\mu_Y^\mathrm{edge})^2 = \frac{ \sqrt{B_0^2-4 A_0 D_0} - B_0}{ 2 A_0}, \quad
	(\mu_N^\mathrm{edge})^2 = 0.
\end{equation}
These expressions will be used to calculate the ``edge variables'' $m_Y^\mathrm{edge}$ and $m_N^\mathrm{edge}$ below.

Making use of the fact that $A_0\geq 0$ and $C_0\geq 0$, we derive the maximum values of allowed $\mu_Y^2$ and $\mu_N^2$ from the inequality~\eqref{eq:ineq} as
\begin{eqnarray}
	(\mu_Y^\mathrm{max})^2	=  \frac{(B_0-C_0)^2}{4 A_0 C_0} - \frac{D_0}{C_0},\quad
	(\mu_N^\mathrm{max})^2 =  \frac{B_0^2}{4 A_0 C_0}- \frac{D_0}{C_0}.
\end{eqnarray}
We can define the ``maximum variables'' as $m_Y^\mathrm{max}\equiv\mu_Y^\mathrm{max}\sqrt{s}$ and $m_N^\mathrm{max}\equiv\mu_N^\mathrm{max}\sqrt{s}$. They are essentially the same as the quantities $\tilde{m}_X^\mathrm{max}$ and $\tilde{m}_N^\mathrm{max}$ originally proposed in Ref.~\cite{HarlandLang:2012gn}. In Fig.~\ref{fig:sketch}, we also demonstrate the point $\big((\mu_Y^\mathrm{max})^2, (\mu_N^\mathrm{max})^2\big)$, which is located slightly beyond the solvable region.

From Fig.~\ref{fig:sketch} we can see that the true mass point would be closer to point $\big((\mu_Y^\mathrm{edge})^2, (\mu_N^\mathrm{edge})^2\big)$ than point $\big((\mu_Y^\mathrm{max})^2, (\mu_N^\mathrm{max})^2\big)$ when the vertex of the parabola is located in quadrant I.
On the other hand, when the vertex is located in quadrant IV, the true mass point may not close to either point, and these variables would not be very useful.

\begin{figure}[!htbp]
\centering
\subfigure[~$e^+e^-\to W^+W^-\to \mu^+\mu^-\nu\bar\nu$\label{fig:1D:ww}]
{\includegraphics[width=.45\textwidth]{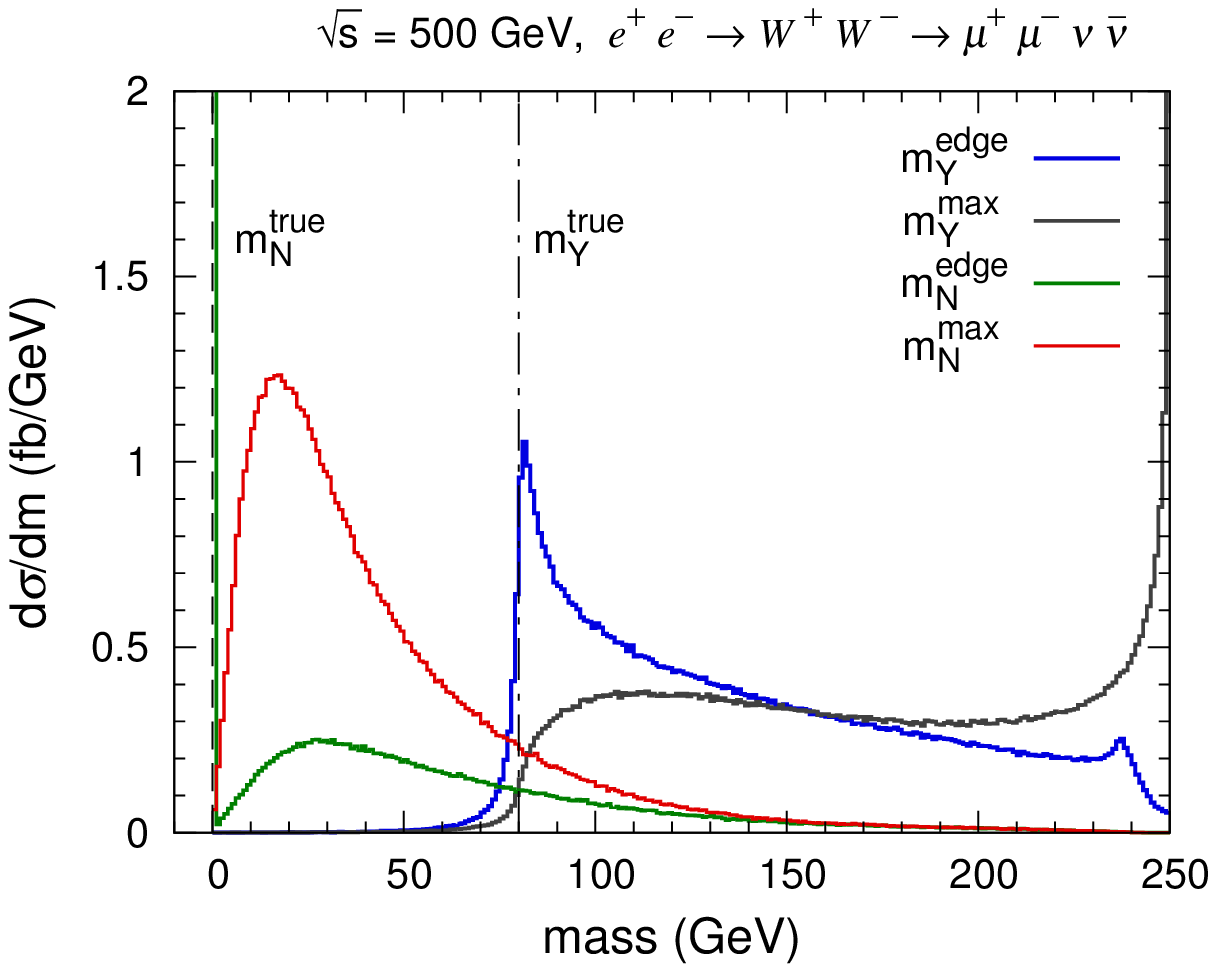}}
\subfigure[~$e^+e^-\to \smur^+\smur^- \to \mu^+\mu^-\tchi_1^0\tchi_1^0$\label{fig:1D:bmp2}]
{\includegraphics[width=.45\textwidth]{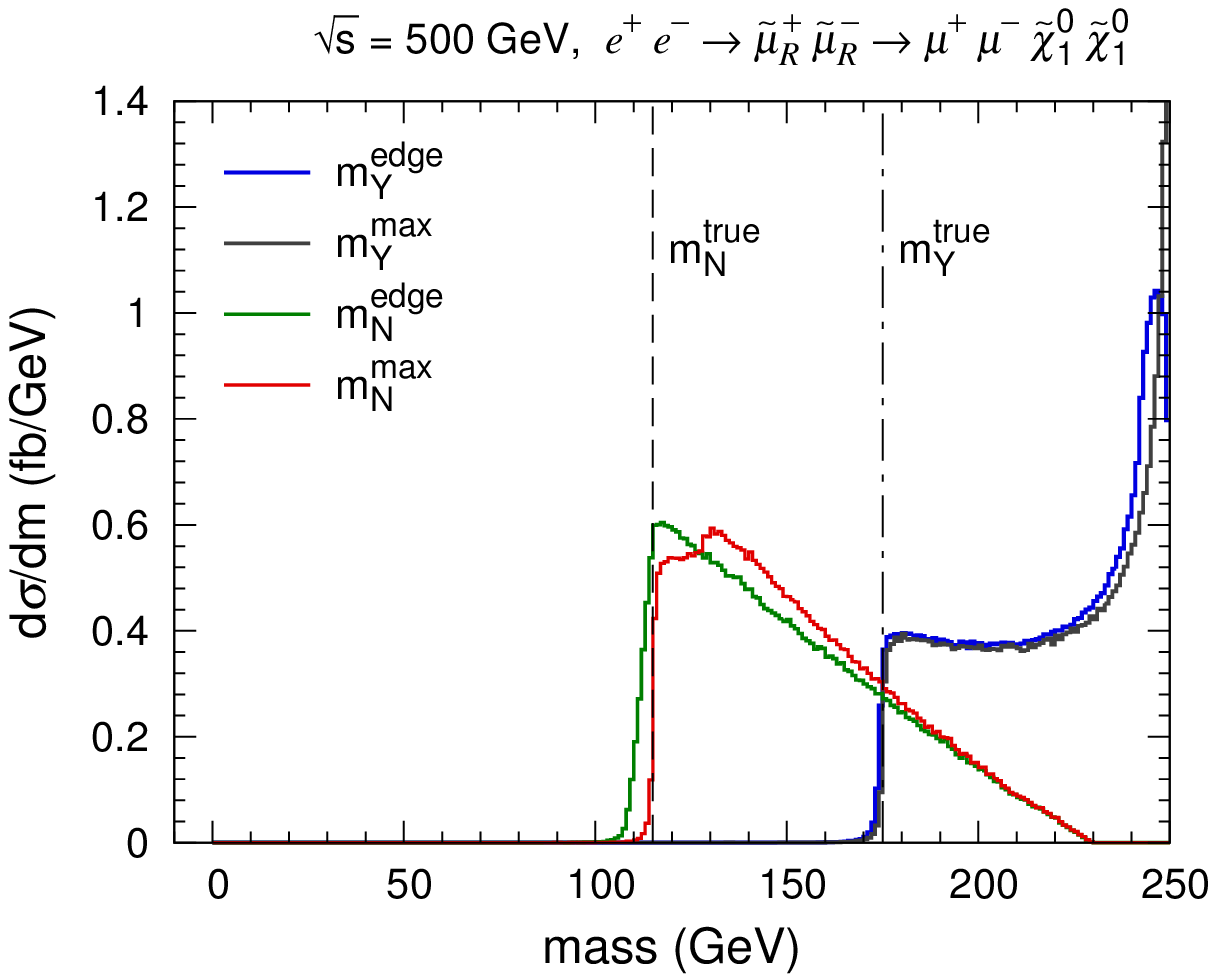}}
\caption{Differential distributions of the edge variables $m_Y^\mathrm{edge}$ and $m_N^\mathrm{edge}$ and the maximum variables $m_Y^\mathrm{max}$ and $m_N^\mathrm{max}$ for $e^+e^-\to W^+W^-\to \mu^+\mu^-\nu\bar\nu$ (a) and $e^+e^-\to \smur^+\smur^- \to \mu^+\mu^-\tchi_1^0\tchi_1^0$ (b).
Dashed (dot-dashed) black lines denote the true mass $m_N^\mathrm{true}$ ($m_Y^\mathrm{true}$).
}
\label{fig:1D}
\end{figure}

We show the differential distributions of $m_Y^\mathrm{edge}$, $m_N^\mathrm{edge}$, $m_Y^\mathrm{max}$, and $m_N^\mathrm{max}$ for $e^+e^-\to W^+W^-\to \mu^+\mu^-\nu\bar\nu$ and $e^+e^-\to \smur^+\smur^- \to \mu^+\mu^-\tchi_1^0\tchi_1^0$ in Figs.~\ref{fig:1D:ww} and \ref{fig:1D:bmp2}, respectively.
The pileup near $\sim 250~\GeV$ in the $m_Y^\mathrm{edge}$ and $m_Y^\mathrm{max}$ distributions are caused by our strategy that takes $\sqrt{s}/2$ as a physical boundary for $m_Y$.
If we extend this boundary to a larger but unphysical value, the distributions will have long tails.
In principle, these variables are bounded from below by the true masses $m_Y^\mathrm{true}$ and  $m_N^\mathrm{true}$.
Consequently, the endpoints of these distributions can be used to extract $m_Y^\mathrm{true}$ and $m_N^\mathrm{true}$.

If $m_N^\mathrm{true}$ is large enough, all distributions would have sharp edges near the true masses, as illustrated in Fig.~\ref{fig:1D:bmp2}.
However, for the case with a small $m_N^\mathrm{true}$, like the SM background $e^+e^-\to W^+W^-\to \mu^+\mu^-\nu\bar\nu$ shown in Fig.~\ref{fig:1D:ww}, the $m_Y^\mathrm{max}$ and $m_N^\mathrm{max}$ distributions would not have sharp edges, while the edges in the $m_Y^\mathrm{edge}$ and $m_N^\mathrm{edge}$ distributions are still quite sharp.
Therefore, in this case the edge variables should be more useful.

\begin{figure}[!htbp]
\centering
\subfigure[~$e^+e^-\to W^+W^-\to \mu^+\mu^-\nu\bar\nu$\label{fig:2Dmax:ww}]
{\includegraphics[width=.45\textwidth]{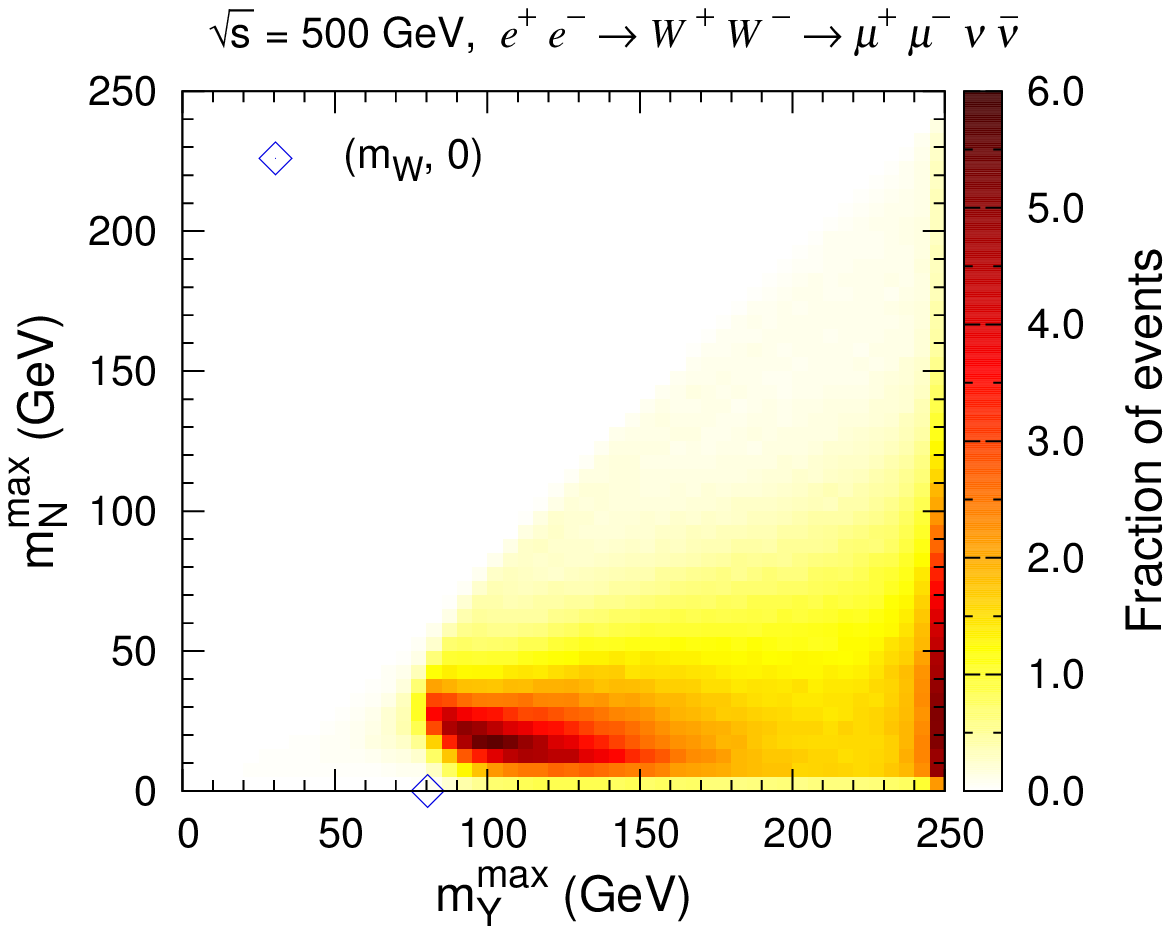}}
\subfigure[~$e^+e^-\to \smur^+\smur^- \to \mu^+\mu^-\tchi_1^0\tchi_1^0$\label{2Dmax:bmp2}]
{\includegraphics[width=.45\textwidth]{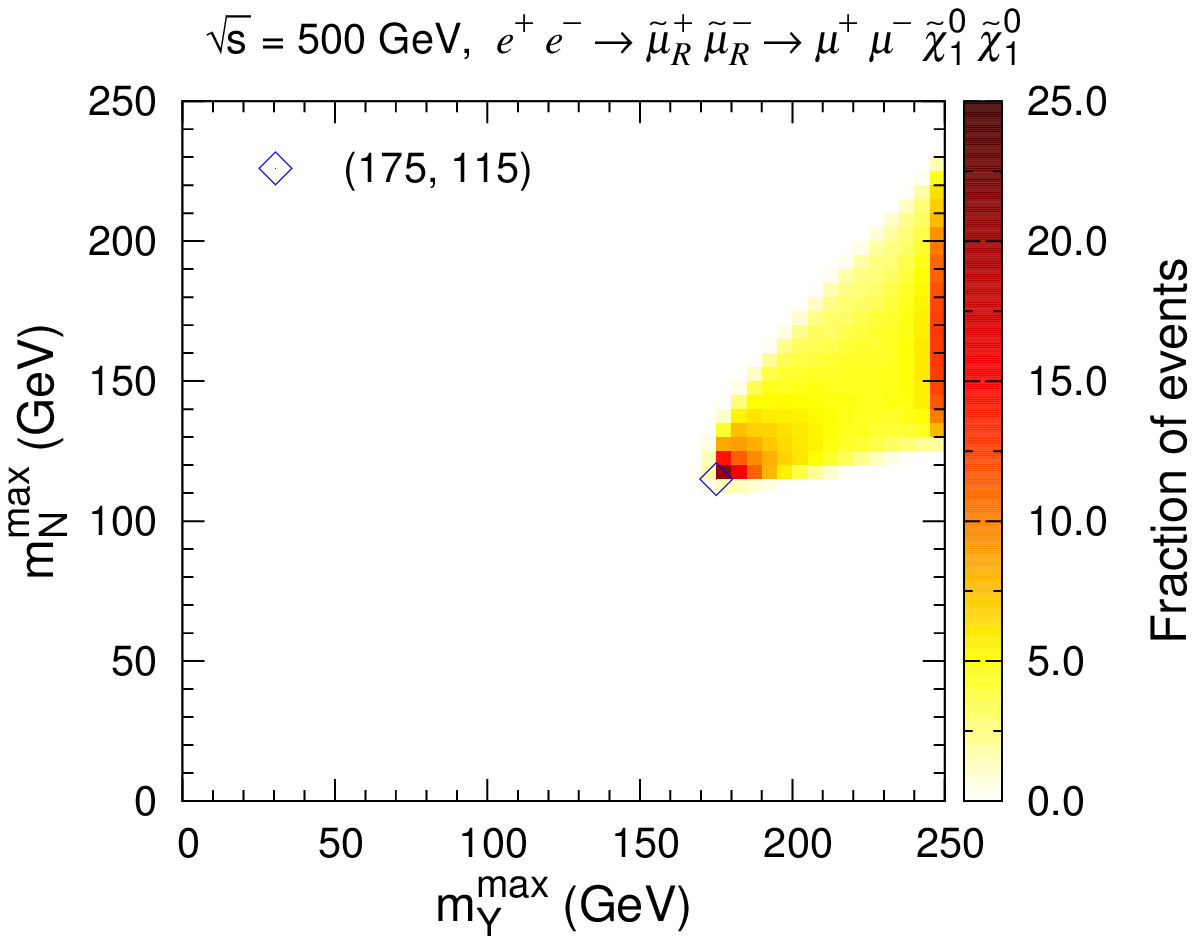}}
\caption{Normalized two-dimensional distributions of on the $m_Y^\mathrm{max}-m_N^\mathrm{max}$ plane for $e^+e^-\to W^+W^-\to \mu^+\mu^-\nu\bar\nu$ (a) and $e^+e^-\to \smur^+\smur^- \to \mu^+\mu^-\tchi_1^0\tchi_1^0$ (b). The empty diamonds denote the true mass points.}
\label{fig:2Dmax}
\end{figure}

In the left (right) panel of Fig.~\ref{fig:2Dmax}, we present the two-dimensional distributions on the $m_Y^\mathrm{max}-m_N^\mathrm{max}$ plane for $e^+e^-\to W^+W^-\to \mu^+\mu^-\nu\bar\nu$ with $m_Y^\mathrm{true}=m_W$ and $m_N^\mathrm{true}=0$ ($e^+e^-\to \smur^+\smur^- \to \mu^+\mu^-\tchi_1^0\tchi_1^0$ with $m_Y^\mathrm{true}=175~\GeV$ and $m_N^\mathrm{true}=115~\GeV$).
While the true mass point in the right panel is located at a very dense region, that point in the left panel is not.
Thus, it is not easy to obtain the true masses from the two-dimensional distribution in the latter case where $m_N^\mathrm{true}$ is small.

\begin{figure}[!htbp]
\centering
\subfigure[~$e^+e^-\to W^+W^-\to \mu^+\mu^-\nu\bar\nu$\label{fig:2Dedge:ww}]
{\includegraphics[width=.45\textwidth]{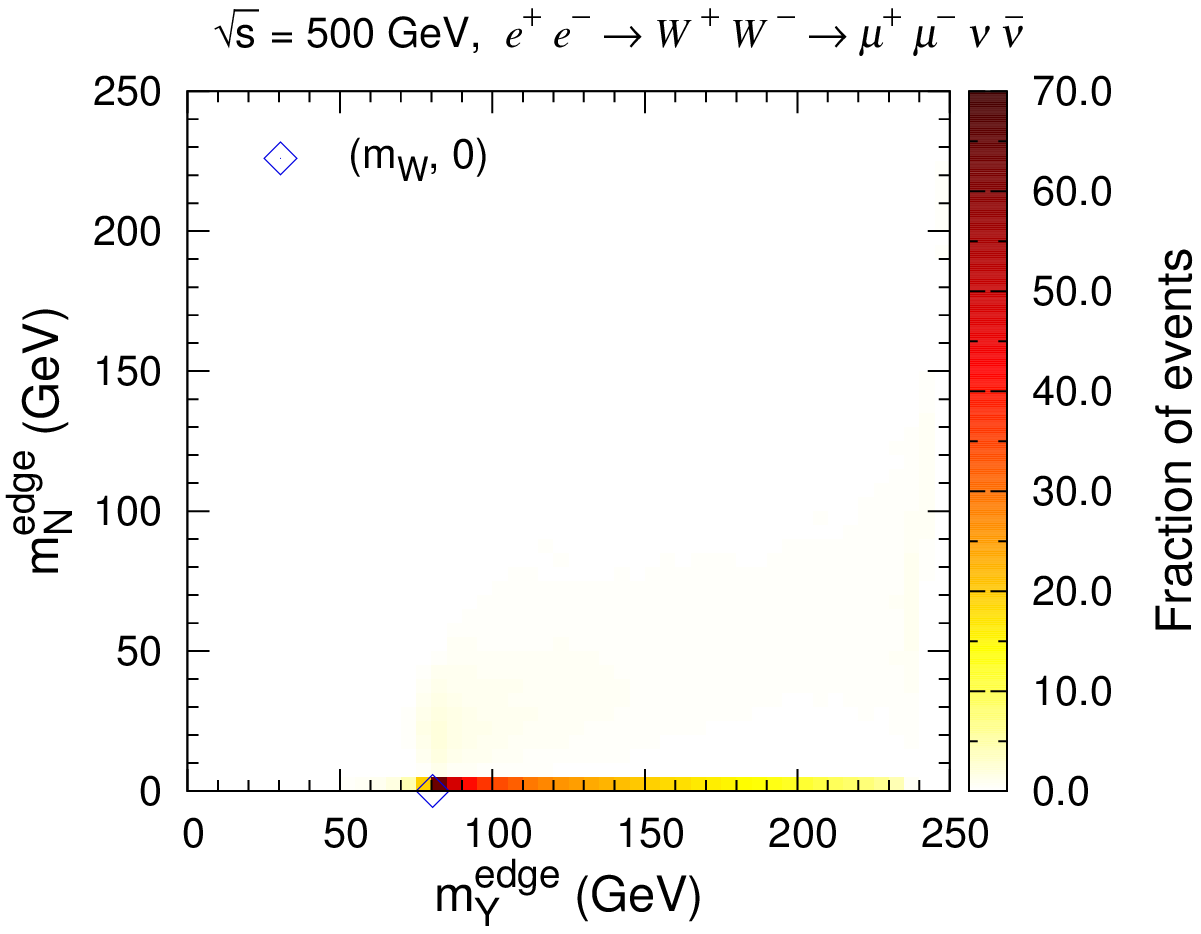}}
\subfigure[~$e^+e^-\to \smur^+\smur^- \to \mu^+\mu^-\tchi_1^0\tchi_1^0$\label{fig:2Dedge:bmp2}]
{\includegraphics[width=.45\textwidth]{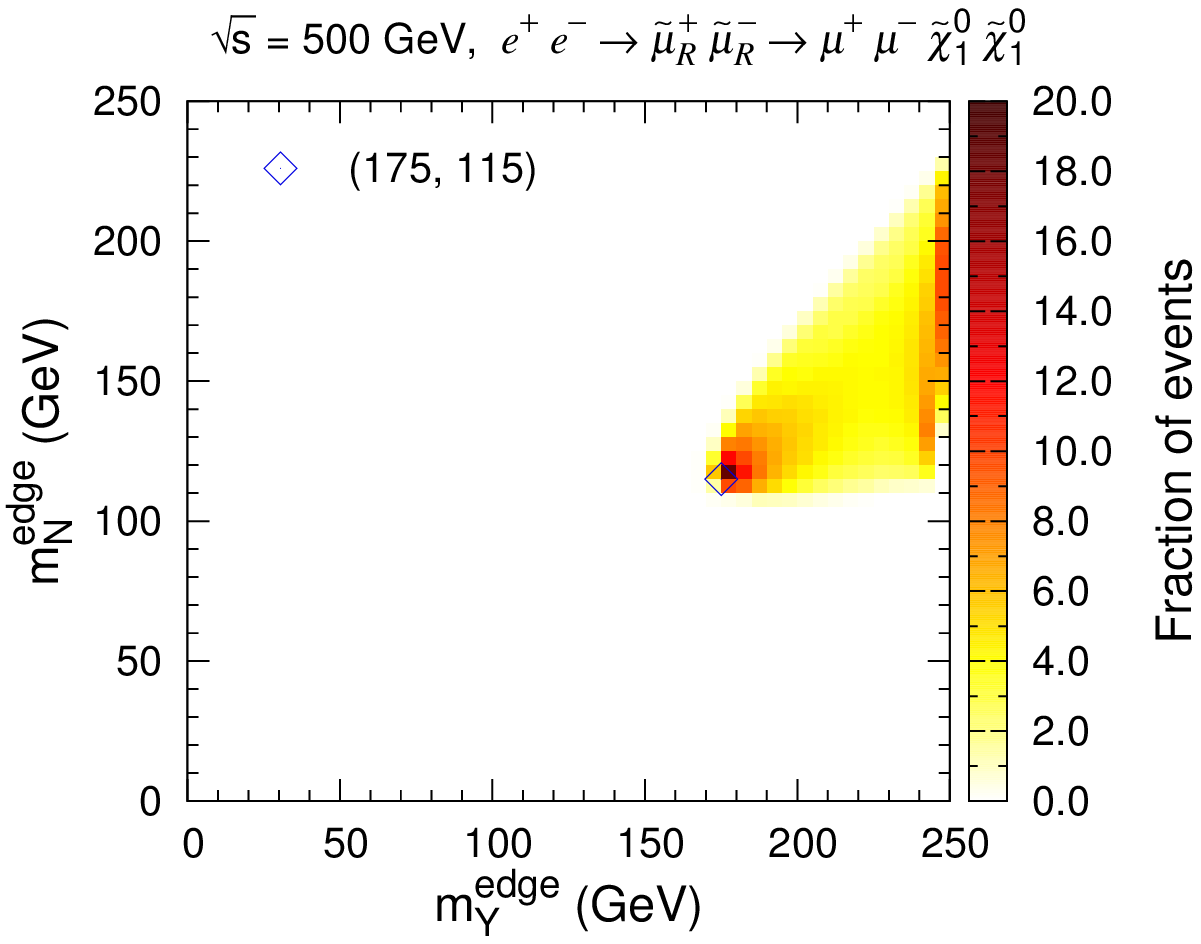}}
\caption{Normalized two-dimensional distributions of on the $m_Y^\mathrm{edge}-m_N^\mathrm{edge}$ plane for $e^+e^-\to W^+W^-\to \mu^+\mu^-\nu\bar\nu$ (a) and $e^+e^-\to \smur^+\smur^- \to \mu^+\mu^-\tchi_1^0\tchi_1^0$ (b). The empty diamonds denote the true mass points.}
\label{fig:2Dedge}
\end{figure}

Then we present the two-dimensional distributions on the $\mY-\mN$ plane in Fig.~\ref{fig:2Dedge}.
Now for both processes, there are distinct peaks corresponding to $(m_Y^\mathrm{true}, m_N^\mathrm{true})$, located at the edges of ($\mY, \mN$) distributions.
This feature can be used to extract the values of $(m_Y^\mathrm{true}, m_N^\mathrm{true})$.
In the next section, we demonstrate how to obtain the true masses from it.

\section{Mass measurement with realistic considerations}
\label{sec:realistic}

In this section, we take into account detector effects, background contamination, and so on. To carry out a fast detection simulation, we adopt \texttt{Delphes~3}~\cite{deFavereau:2013fsa} with a setup for the International Large Detector.

\subsection{Selection cuts}

The process $e^+e^-\to \smur^+\smur^- \to \mu^+\mu^-\tchi_1^0\tchi_1^0$ is considered as the signal.
Its leads to the $\mu^+\mu^- + \missp$ final state at $e^+e^-$ colliders, where $\missp$ denotes the missing momentum due to the invisible particles.
In order to illustrate the efficiency of the selection cuts and the mass extraction method, for an $e^+e^-$ collider with $\sqrt{s}=500~\GeV$ we choose three benchmark points:
\begin{itemize}
\item \textbf{BP1:} $m_{\smur}=135~\GeV$, $m_{\tchi_1^0}=45~\GeV$;
\item \textbf{BP2:} $m_{\smur}=175~\GeV$, $m_{\tchi_1^0}=115~\GeV$;
\item \textbf{BP3:} $m_{\smur}=175~\GeV$, $m_{\tchi_1^0}=155~\GeV$.
\end{itemize}
The following selection cuts are adopted to efficiently suppress backgrounds.
\begin{itemize}
	\item \textbf{Lepton cut:}
	select the events with exactly two opposite-sign muons with $\pT > 10~\GeV$ and $|\eta| < 2.4$;
	veto the events containing any electron with $\pT > 10~\GeV$ and $|\eta| < 2.4$.

	\item \textbf{$\missET $ cut:}
	select the events with $\missET > 5~\GeV$.

	\item \textbf{$\Delta \phi$ cut:}
	select the events with $\Delta \phi (\mu^+,\mu^-) < 2.4$.

	\item \textbf{$m_{\mu\mu}$ cut:}
	reject the events with $|m_{\mu\mu} - m_Z| < 10~\GeV$, where $m_{\mu\mu}$ is the invariant mass of the two muons;
	reject the events with $m_{\mu\mu} > 220~\GeV$ or $m_{\mu\mu} < 10~\GeV$.
\end{itemize}

The irreducible SM background is the 4-body production $e^+e^- \to \mu^+ \mu^- \nu\bar{\nu}$, which mainly comes from $W^+W^-$ and $ZZ$ production.
The $W^+W^-$ production $e^+e^- \to W^+ W^- \to \mu^+ \nu_\mu \mu^- \bar{\nu}_\mu$ is dominant and larger than the $ZZ$ production $e^+e^- \to ZZ \to \mu^+\mu^- \bar{\nu}_\ell\nu_\ell$ by an order of magnitude.
There are many other diagrams without two $s$-channel massive vector bosons; their contributions cannot be neglected.
We directly generate the $e^+e^- \to \mu^+ \mu^- \nu\bar{\nu}$ background sample taking into account all the diagrams and interference effects.

Minor backgrounds include $e^+e^- \to \mu^+ \mu^-$ and $e^+e^- \to \tau^+ \tau^-$ where the taus subsequently decay to muons with a branching ratio of $17.4\%$.
The cut on $\missET$ is very helpful for suppressing the $\mu^+\mu^-$ background, because there is no genuine missing momentum source in it.
Furthermore, since the tau mass is negligible compared with $\sqrt{s}$, the tau pairs produced are highly boosted.
As a result, the two muons in the final state, either directly produced or from tau decays, will be back to back with large $\Delta \phi (\mu^+, \mu^-)$.
The requirement of $\Delta \phi(\mu^+, \mu^-) < 2.4$ is useful for suppressing these two backgrounds.
It is observed that after the $\missET$ and $\Delta \phi(\mu^+, \mu^-)$ cuts they become negligible.

\begin{figure}[!htbp]
\centering
\includegraphics[width=0.45\textwidth]{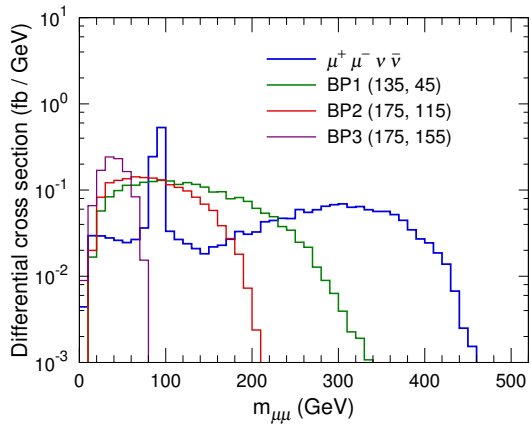}
\caption{$m_{\mu\mu}$ distributions for the background and the signal benchmark points at $\sqrt{s} = 500~\GeV$.}
\label{fig:m2mu}
\end{figure}

In Fig.~\ref{fig:m2mu}, we show the differential cross sections and normalized distributions of $m_{\mu\mu}$ for the $e^+e^- \to \mu^+ \mu^- \nu\bar{\nu}$ background and the signals from the three benchmark points.
For suppressing the background, we reject the events with $m_{\mu \mu} > 220$ GeV.
This cut has less influence on the signal, and we use a fixed threshold for all benchmark points for simplicity.
Moreover, the events with $m_{\mu\mu} < 10~\GeV$ are vetoed in order to remove events from quarkonium decays. Additionally, there is a distinct peak around the $Z$ pole in the background; it is remove by rejecting the events in the window $|m_{\mu\mu} - m_Z| < 10~\GeV$.


\begin{table}[!htb]
\centering
\setlength\tabcolsep{.4em}
\renewcommand{\arraystretch}{1.3}
\begin{tabular}{ccccccccc}
\hline
\hline
                     & $\mu^+ \mu^- \nu\bar{\nu}$ & $W^+W^-$ & $ZZ$ & $\mu^+ \mu^-$ & $\tau^+ \tau^-$ & BP1   & BP2  & BP3 \\\hline
No cut	& 96.08   & 61.79   & 4.89 & 419.70        & 419.91          & 59.64 &36.05 &36.05 \\
Lepton cut        & 65.71   &47.50 &  3.73  & 369.67        & 9.99            & 47.29 &28.42 &20.23 \\
$\missET$ cut     & 64.83   & 46.66 & 3.72   &9.57           & 9.41            & 47.09 &28.18 &19.18 \\
$\Delta \phi$ cut & 23.55   & 12.21 & 2.70   &$\sim 0$       &$\sim 0$         & 22.00 &16.65 &9.81  \\
$m_{\mu\mu}$  cut & 5.34    & 2.76 & 0.18   &$\sim 0$       &$\sim 0$         & 17.38 &14.01 &9.72 \\
Cut efficiency (\%)       & 5.55    &  -   & -      & -             &-                & 29.15 &38.86 &26.95\\
\hline
\hline
\end{tabular}
\caption{Visible cross sections $\sigma$ (in fb) for backgrounds and signal benchmark points after each cut at the detector level. Note that the $\mu^+ \mu^- \nu\bar{\nu}$ background (the second column) actually includes the $W^+W^- \to \mu^+ \mu^- \nu_\mu\bar{\nu}_\mu$ (the third column) and $ZZ \to \mu^+\mu^- \bar{\nu}_\ell\nu_\ell$ (the fourth column) backgrounds, which are listed here to specify the on-shell diboson contributions.}
\label{tab:xsec}
\end{table}

In Table.~\ref{tab:xsec}, we list the visible cross sections for the backgrounds and the signal benchmark points in each stage.
The production cross sections of the last two benchmark points are equal because they correspond to the same smuon mass and the phase space with off-shell smuons is negligible.
It is obvious that the cut conditions we adopted are efficient.
After imposing all the cuts, the cross sections of SM backgrounds are smaller than that of the signals, but they cannot be neglected.
It might be possible to further reduce the backgrounds with some sophisticated cuts.
However, the current event selections should be adequate for the mass measurement we are going to discuss.

\subsection{Mass measurement}
\label{sec:mass_meas}

\begin{figure}[!htbp]
\centering
\subfigure[~Without the cuts]{
\includegraphics[width=0.45\textwidth]{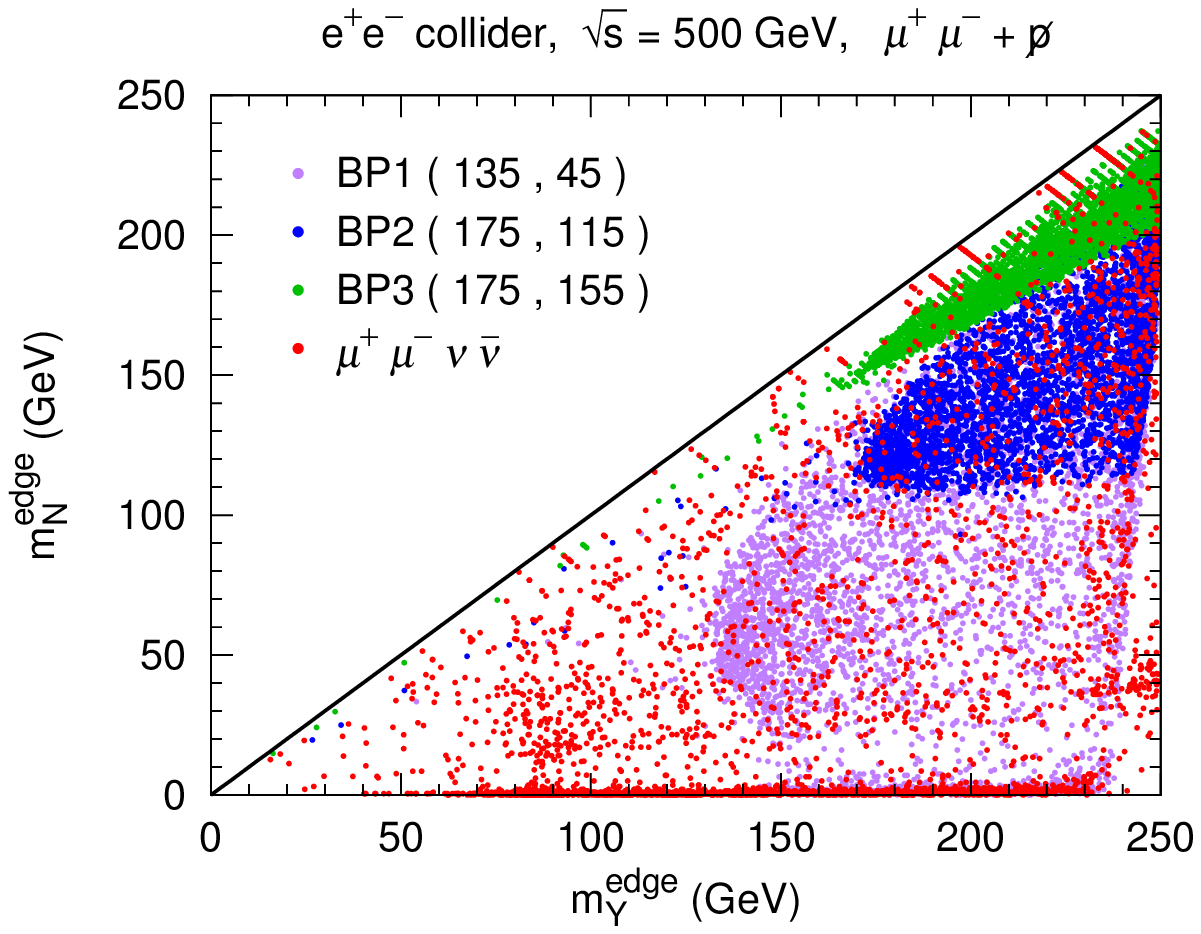}}
\subfigure[~After the cuts]{
\includegraphics[width=0.45\textwidth]{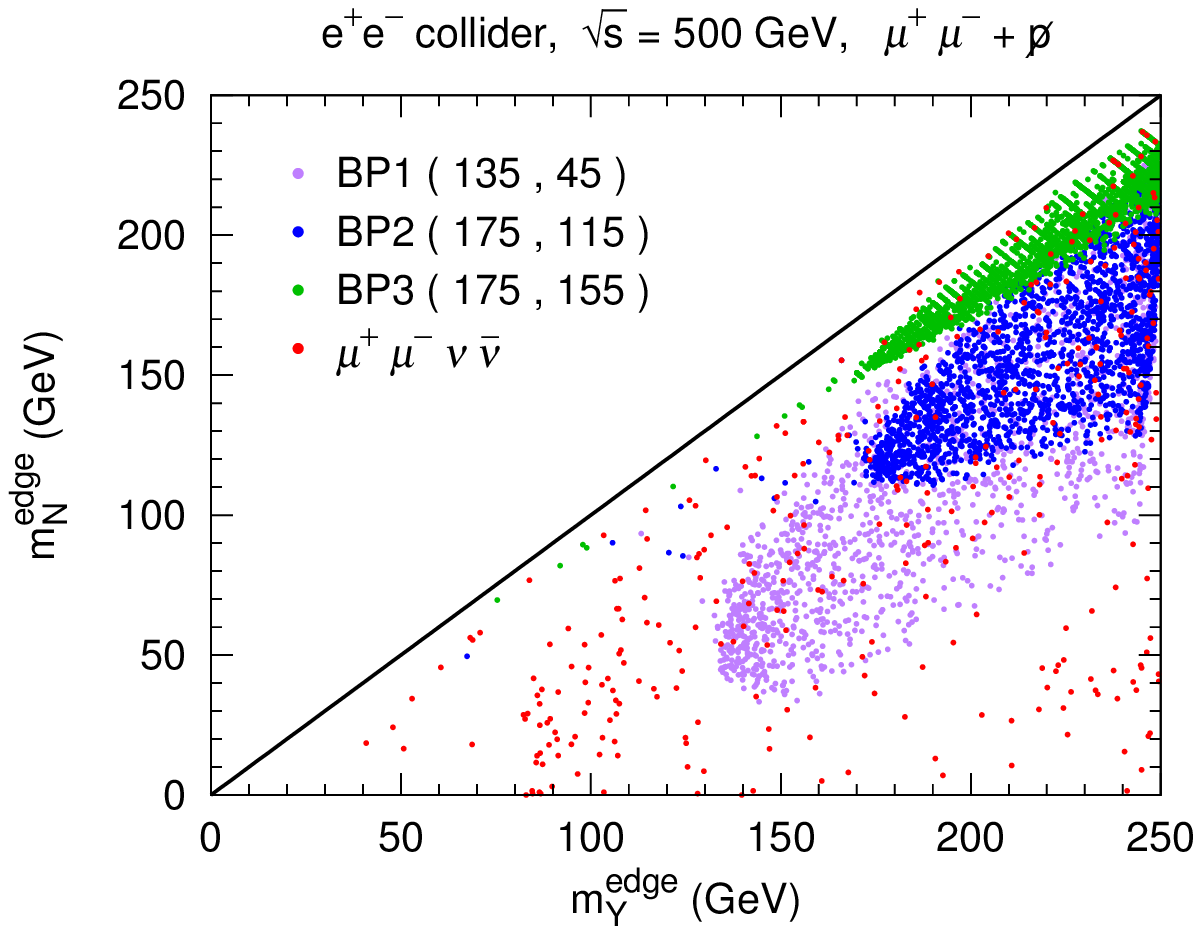}}
\caption{Scatter plots of the simulation events on the $\mY$-$\mN$ plane for the $\mu^+ \mu^- \nu\bar{\nu}$ background and the three benchmark points before (a) and after (b) the cuts. It is assumed that there are 5,000 events for each process before the cuts.}
\label{fig:dist}
\end{figure}

We show the scatter plots on the $\mY$-$\mN$ plane for the backgrounds and the three benchmark points in Fig.~\ref{fig:dist}, assuming 5,000 events for each process before the cuts and taking into account the detector effects.
As can be seen, the distribution for the $\mu^+ \mu^- \nu\bar{\nu}$ background spreads in the whole triangle region with $\mY > \mN$, while the distributions of the signals are bounded by $m_Y^\mathrm{true}$ and $m_N^\mathrm{true}$.
Without the cuts, there are a lot of background events clustering around the $\mN = 0$ line with $\mY > m_W$;
most of them come from the on-shell $W^+W^-$ production due to the same reason for the behavior in Fig.~\ref{fig:2Dedge:ww}, and can be removed by the $m_{\mu\mu}$ cuts.
After imposing all the selection cuts, the background is efficiently reduced.

\begin{figure}[!htbp]
\centering
\subfigure[~Background + BP1]{
\includegraphics[width=0.333\textwidth]{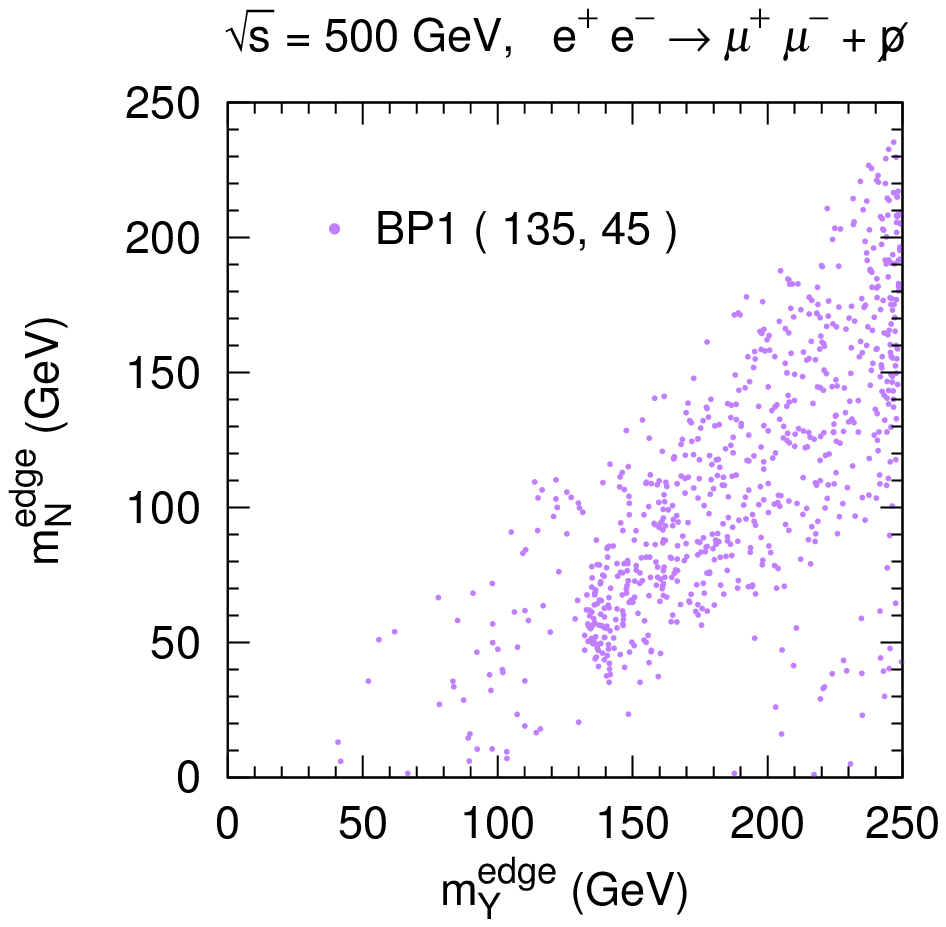}}%
\subfigure[~Background + BP2]{
\includegraphics[width=0.333\textwidth]{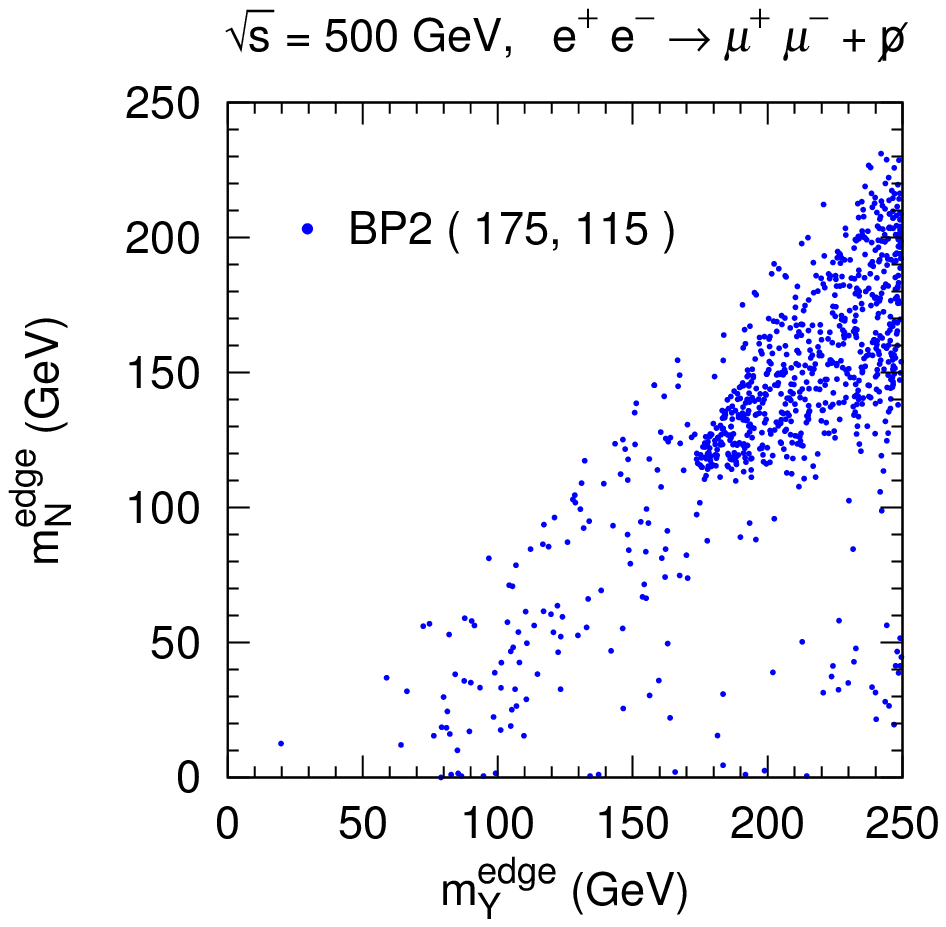}}%
\subfigure[~Background + BP3]{
\includegraphics[width=0.333\textwidth]{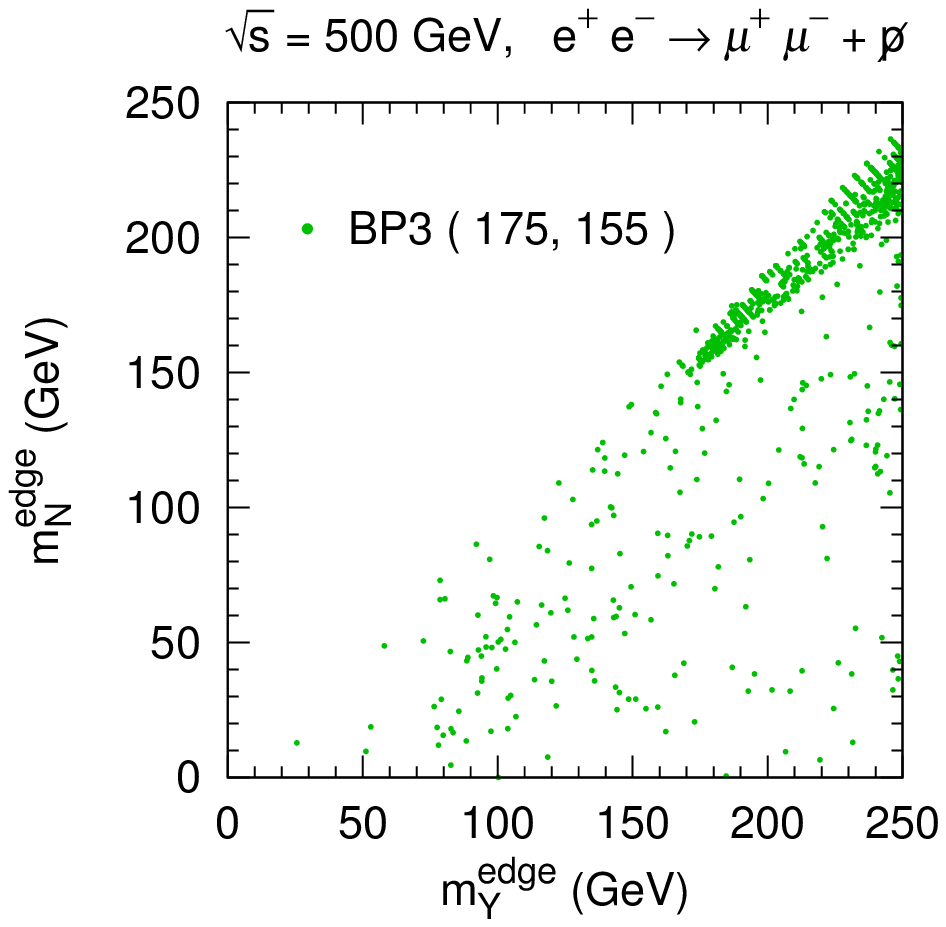}}
\caption{Scatter plots on the $\mY$-$\mN$ plane for the events generated by simultaneously simulating a signal benchmark point and the background  with the selection cuts applied. It is assumed that there are 5,000 events before the cuts.}
\label{fig:dist_s}
\end{figure}

Since it is difficult to further distinguish the signals from the background, the contamination would be unavoidable.
Therefore, we generate the events simultaneously induced by the background and each benchmark point.
In Fig.~\ref{fig:dist_s}, we present the scatter plot for these events.
The boundary of the signal is still quite clear, despite the background contamination.

Below we attempt to extract the physical values of $m_Y$ and $m_N$ from the $(\mY,\mN)$ distributions.
It is expected that the events around the true masses should be very dense.
This is demonstrated in Fig.~\ref{fig:2Dedge}, where we show the fraction of events in each grid on the $\mY$-$\mN$ plane for the benchmark points.
Note that many events locate in a region near $\mY \sim \sqrt{s}/2$, due to the physical boundary we use in the algorithm to derive $\mY$ and $\mN$, as mentioned above for explaining the behavior in Fig.~\ref{fig:1D}.
In order to avoid the disturbance from these events, we exclude the events with $\mY < 240~\GeV$, and hence the event density around the true masses is expected to be the highest.

\begin{figure}[!htbp]
\centering
\subfigure[~Background + BP1]{
\includegraphics[width=0.333\textwidth]{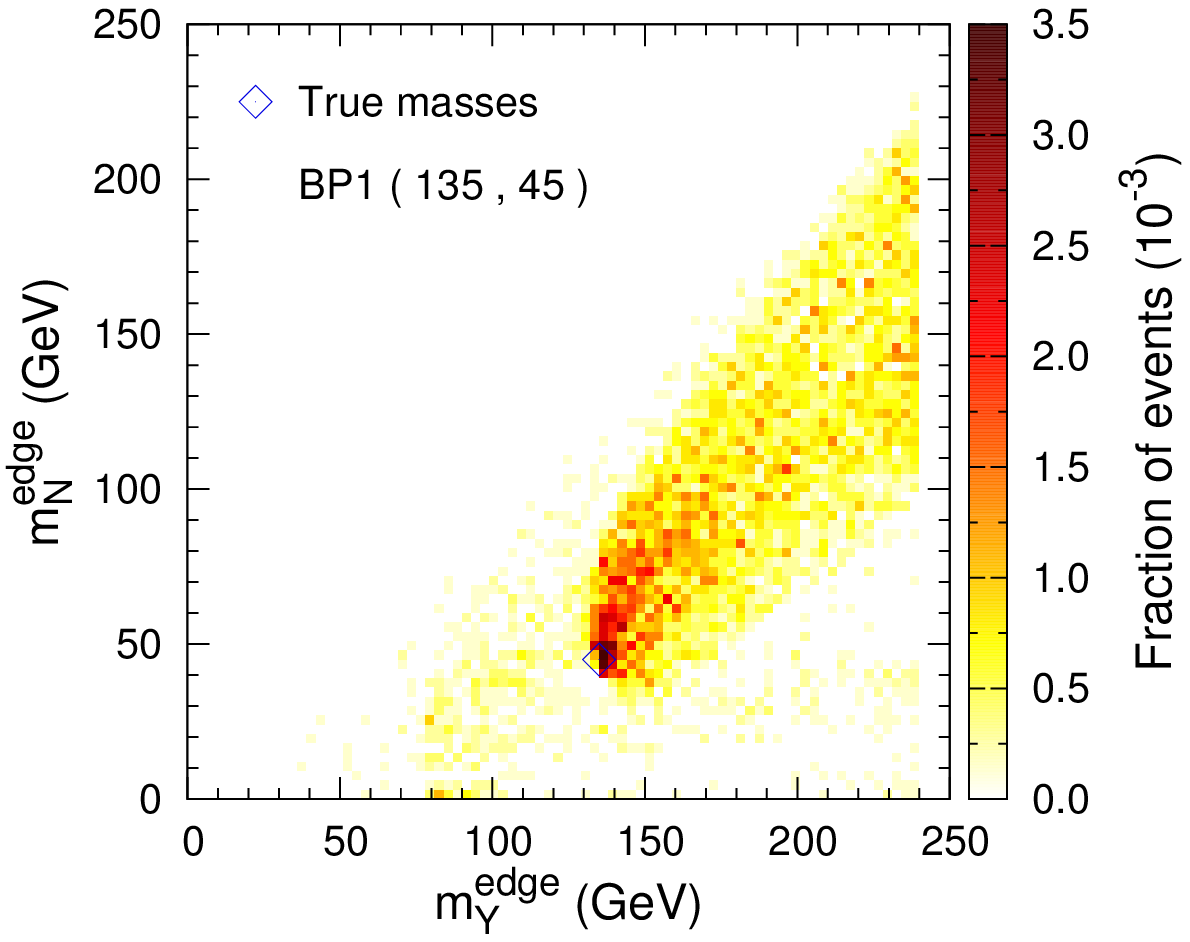}}%
\subfigure[~Background + BP2]{
\includegraphics[width=0.333\textwidth]{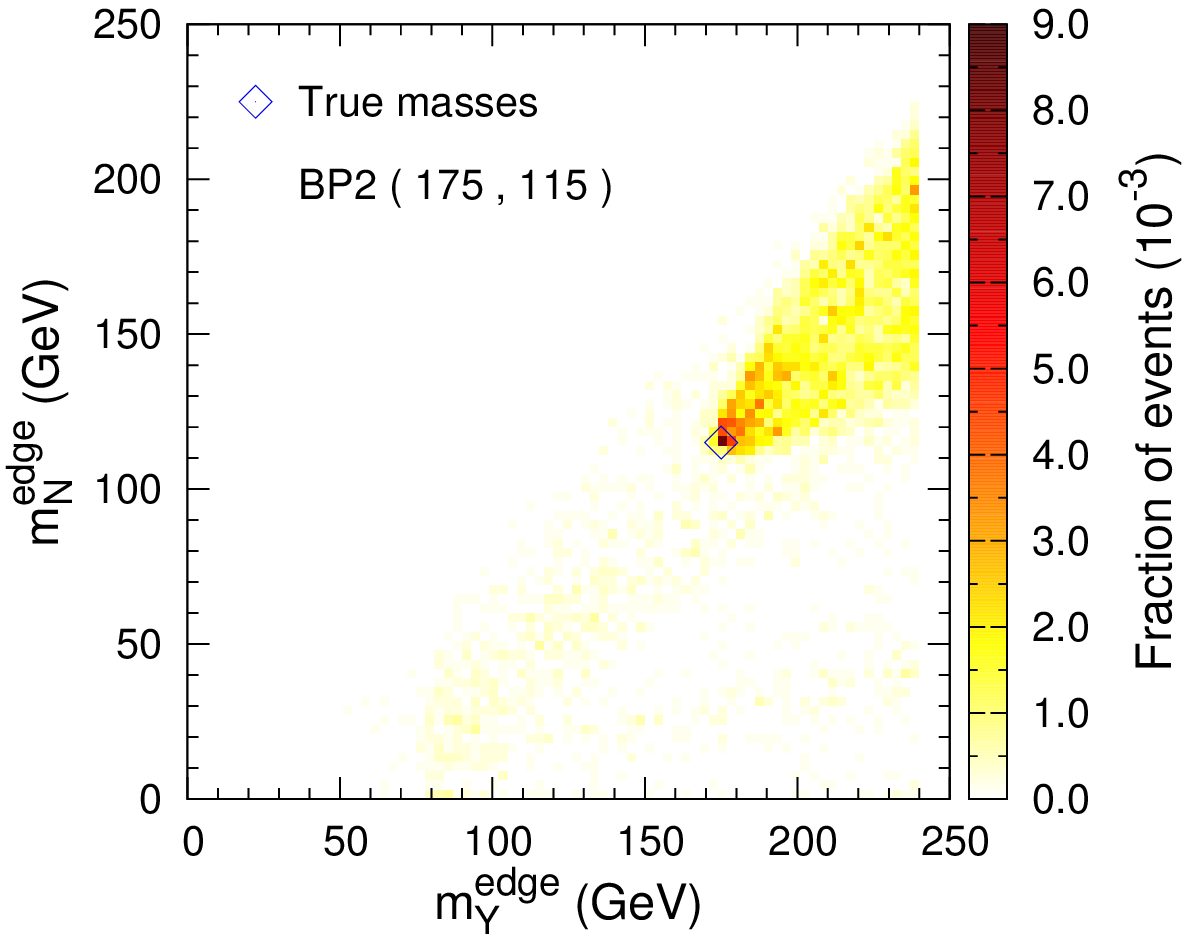}}%
\subfigure[~Background + BP3]{
\includegraphics[width=0.333\textwidth]{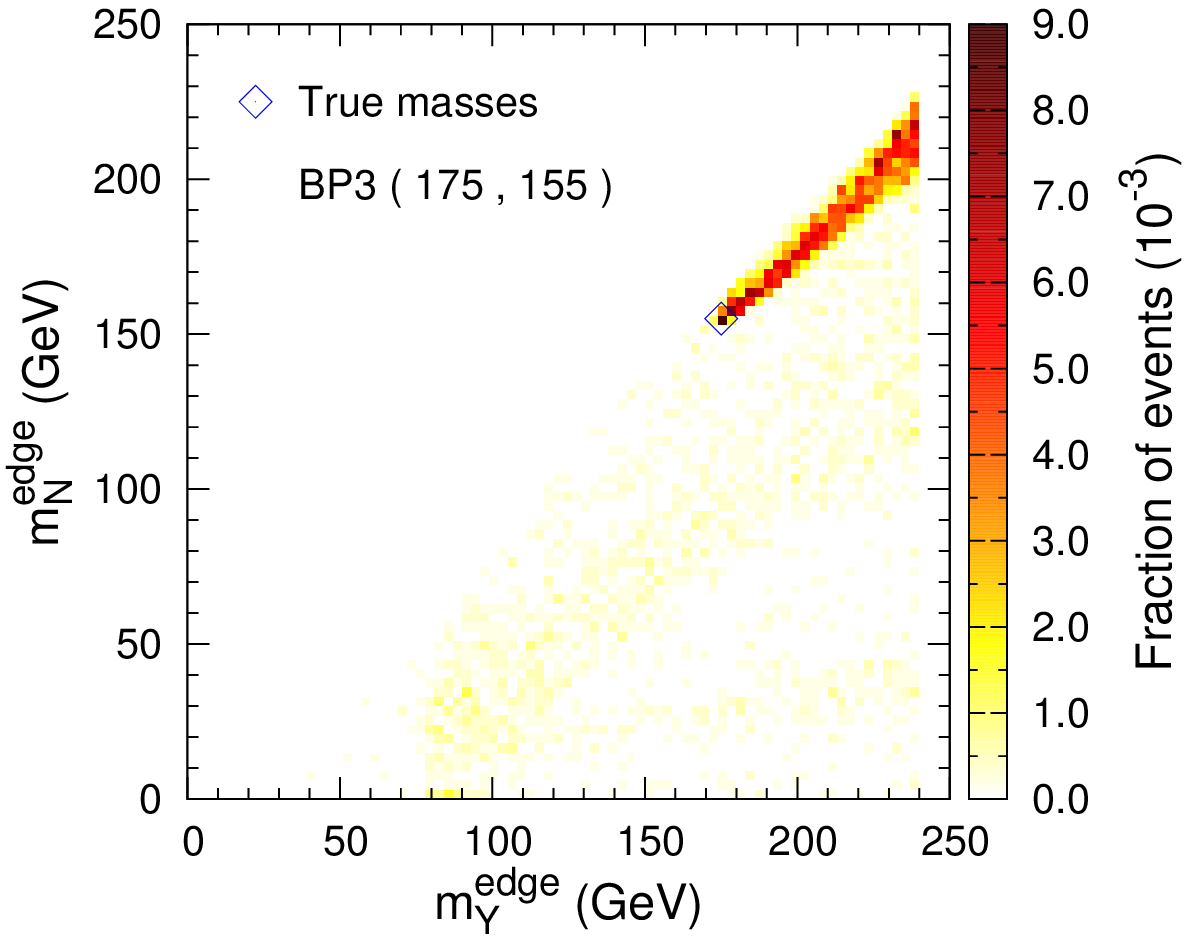}}
\caption{Normalized two-dimensional distributions on the $\mY$-$\mN$ plane for the events from three signal benchmark points contaminated by the background.
The empty diamonds denote the true values of $(m_Y, m_N)$.}
\label{fig:2Ds}
\end{figure}

\begin{figure}[!htbp]
\centering
\includegraphics[width=0.3\textwidth]{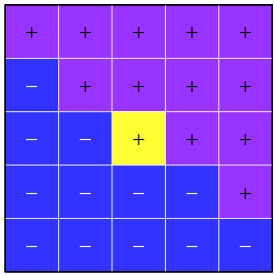}
\caption{Illustration of the definition of the refined density for the yellow grid. The plus and minus signs represent the addition and subtraction of the event density, respectively.}
\label{fig:enh}
\end{figure}

\begin{figure}[!htbp]
\centering
\subfigure[~Background + BP1]{
\includegraphics[width=0.333\textwidth]{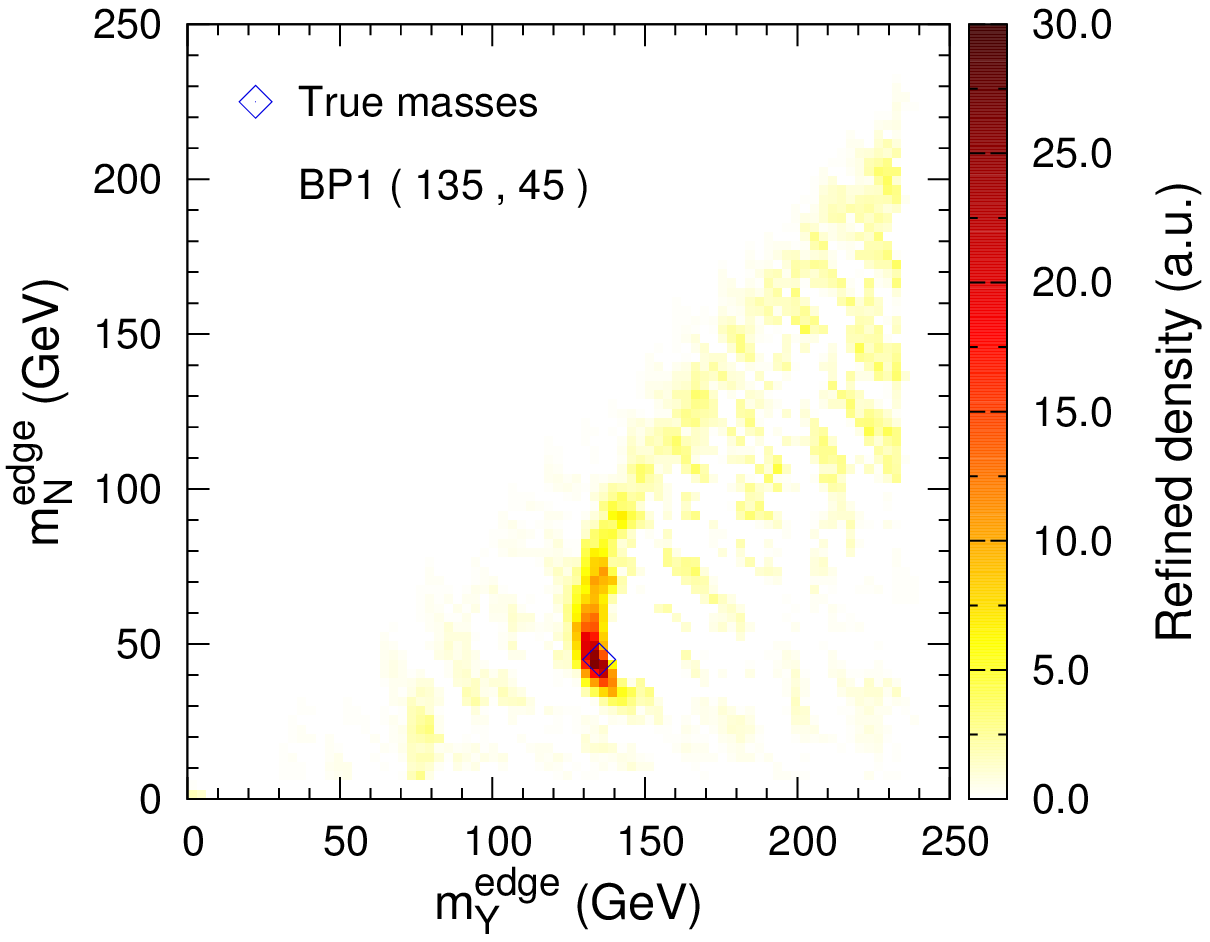}}%
\subfigure[~Background + BP2]{
\includegraphics[width=0.333\textwidth]{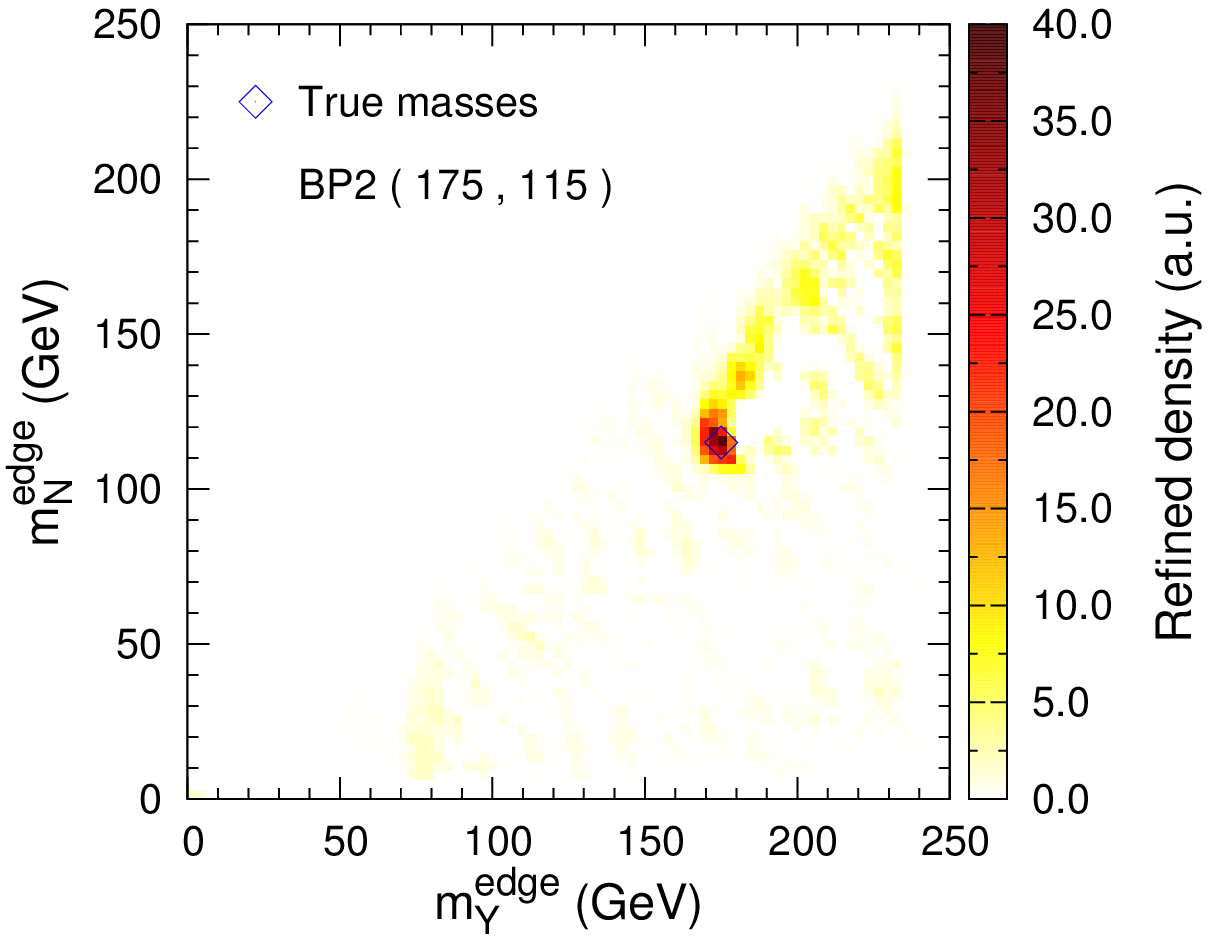}}%
\subfigure[~Background + BP3]{
\includegraphics[width=0.333\textwidth]{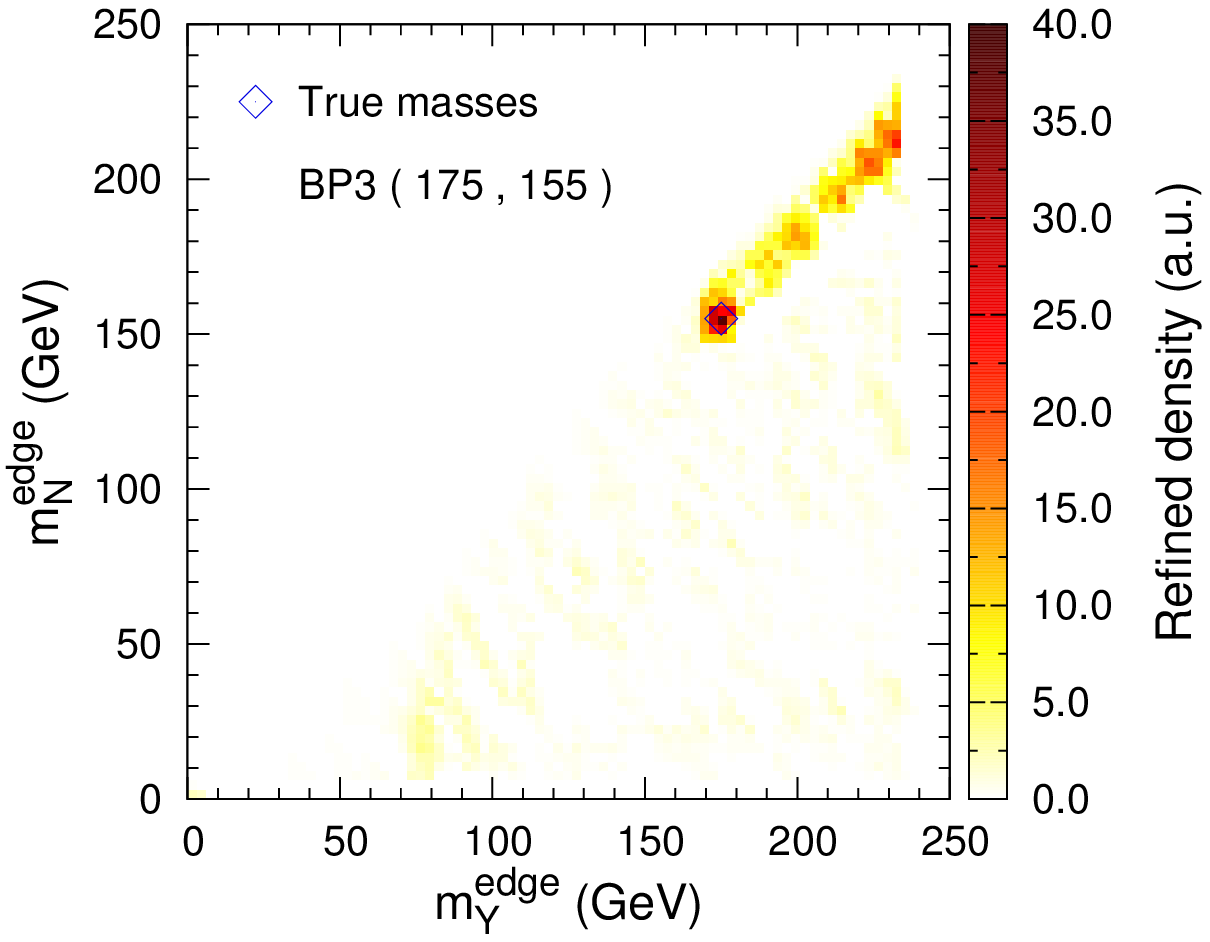}}
\caption{2-dimensional distributions of the refined density (arbitrary unit) on the $\mY$-$\mN$ plane for three signal benchmark points contaminated by the background.
The empty diamonds denote the true values of $(m_Y, m_N)$.}
\label{fig:enh_s}
\end{figure}

However, due to the fluctuation of the events, the above argument is not always true, particularly when we shrink the size of the grids.
In order to avoid this issue, we could also utilize the information of the grids neighbored to each grid.
In Fig.~\ref{fig:2Ds}, it is observed that the upper right grids neighbored to the grid enclosing the true mass point are also quite dense while the lower left grids are very sparse.
Therefore, we make use of this feature and calculate the refined density, to which the upper right grids have positive contributions but the lower left grids have negative contributions.
In practice, as demonstrated in Fig.~\ref{fig:enh}, we consider the 24 grids around each grid.
The refined density for the yellow grid is defined as the sum of the event densities of the yellow and purple grids, subtracted by the event densities of the blue grids.
Fig.~\ref{fig:enh_s} shows the distributions of this refined density, where the true mass point is much more probable to lie within the densest grid.

Then we can treat the center of the densest grid as the measured masses. Apparently, the measurement precision is controlled by the grid size. In order to increase the precision, we should shrink the grid size as small as possible, as long as the result is stable.
The details are described in Appendix~\ref{sec:ext}, where the stability is judged by an inequality.
We iterate the procedure until this inequality is violated.
Finally, the measured values $(m_Y^\mathrm{meas}, m_N^\mathrm{meas})$ are obtained.

\begin{table}[!htb]
\centering
\setlength\tabcolsep{.4em}
\renewcommand{\arraystretch}{1.3}

Edge Variables\\[.2em]
\begin{tabular}{ccccccc}
\hline
\hline
  & \multicolumn{2}{c}{BP1}  & \multicolumn{2}{c}{BP2} & \multicolumn{2}{c}{BP3}\\
  Processes  & $m_Y$ & $m_N$  & $m_Y$ & $m_N$  & $m_Y$ & $m_N$ \\
\hline
True masses  &135  &45   &175 & 115  & 175 & 155 \\
PL no bkg, $500~\ifb$ &$134.4 \pm 0.8$ &$43.4 \pm 1.5$  &$175.2 \pm 0.3$& $114.2 \pm 0.3$
               &$174.9 \pm 0.3$ &$155.1 \pm 0.4$ \\
DL no bkg, $500~\ifb$ &$134.1 \pm 0.7$ &$43.5 \pm 1.4$  &$175.1 \pm 0.4$& $114.2 \pm 1.0$
               &$174.8 \pm 0.6$ &$154.9 \pm 0.5$ \\
DL + bkg, $100~\ifb$	&$134.0 \pm 1.2$ &$45.2 \pm 2.6$  &$174.4 \pm 1.2$& $116.2 \pm 1.8$
&$176.8 \pm 6.6$ &$156.7 \pm 6.7$ \\
DL + bkg, $500~\ifb$	&$134.4 \pm 0.7$ &$44.3 \pm 1.7$  &$175.1 \pm 0.6$& $114.8 \pm 1.2$
               &$174.9 \pm 0.5$ &$155.3 \pm 0.6$ \\
DL + bkg, $1000~\ifb$ 	&$134.7 \pm 0.5$ &$43.7 \pm 1.2$  &$175.1 \pm 0.4$& $114.6 \pm 1.0$
               &$174.8 \pm 0.5$ &$155.2 \pm 0.6$ \\
\hline
\hline
\end{tabular}

\vspace{1em}
Maximum variables\\[.4em]
\begin{tabular}{ccccccc}
\hline
\hline
   & \multicolumn{2}{c}{BP1}  & \multicolumn{2}{c}{BP2} & \multicolumn{2}{c}{BP3}\\
  Processes & $m_Y$ & $m_N$  & $m_Y$ & $m_N$  & $m_Y$ & $m_N$ \\
\hline
DL + bkg, $100~\ifb$	&$136.1 \pm 1.3$ &$47.9 \pm 2.4$  &$175.6 \pm 1.1$& $116.9 \pm 1.9$
               &$180.0 \pm 13.8$ &$159.3 \pm 13.5$ \\
DL + bkg, $500~\ifb$	&$135.8 \pm 1.1$ &$48.0 \pm 1.7$  &$175.5 \pm 0.6$& $116.7 \pm 1.5$
               &$175.1 \pm 0.5$ &$155.5 \pm 0.6$ \\
DL + bkg, $1000~\ifb$ 	&$135.9 \pm 0.6$ &$48.0 \pm 1.2$  &$175.5 \pm 0.5$& $116.2 \pm 1.5$
               &$175.0 \pm 0.5$ &$155.4 \pm 0.6$ \\
\hline
\hline
\end{tabular}
\caption{Mean values and uncertainties of the measured masses for three benchmark points from different processing approaches  with specified integrated luminosities using edge and maximum variables.
``PL'' and ``DL'' denote the results obtained from parton-level simulation and detector-level simulation, respectively.
``+ (no) bkg'' indicates that background contamination is (not) taken into account.
}
\label{tab:bmp}
\end{table}

In order to estimate the uncertainty in the mass measurement, for each benchmark point we perform several simulations and derive the masses of new particles. The mean values of the measured masses in these pseudo-experiments are taken to be $m_Y^\mathrm{meas}$ and $m_N^\mathrm{meas}$ hereafter. Along with the mean values, we also calculate the uncertainties of the measured masses, $\sigma_{m_Y}$ and $\sigma_{m_N}$. For three benchmark points with an integrated luminosity of $500~\ifb$, the mean values and the uncertainties of the measured masses are listed in Table~\ref{tab:bmp}.

In Table~\ref{tab:bmp}, we also show the impacts of detector effects and SM background contamination. Compared with the results from simulations at parton level, the simplified detector effects would not significantly affect the mean values and uncertainties. If the background contamination is taken into account, the uncertainties become slightly larger. Finally, we calculate the results for integrated luminosities of $100~\ifb$ and $1000~\ifb$. We find that larger integrated luminosities can reduce the uncertainties.

For comparison, we also list the results given by the maximum variables $m_Y^\mathrm{max}$ and $m_N^\mathrm{max}$ in Table.~\ref{tab:bmp}.
For BP2 and BP3, the precision of the mass measurement using the edge variables is comparable to or slightly better than that using the maximum variables. Moreover, for BP1 with a small $m_N^\mathrm{true}$, the result of the $m_N$ measurement using the edge variables is much better.

\subsection{Measurement precision}

In this subsection, we study the measurement precision of our method in general.
We consider a setup for an $e^+e^-$ collider at $\sqrt{s}=500~\GeV$ with a data set of $500~\ifb$ collected and generate simulation samples for various values of $m_Y^\mathrm{true}$ and $m_N^\mathrm{true}$ ranging from 5 to $245~\GeV$ with a step size of $10~\GeV$.
The measurement deviations $|m_Y^\mathrm{meas}-m_Y^\mathrm{true}|$ and $|m_N^\mathrm{meas}-m_N^\mathrm{true}|$, as well as the uncertainties of the measured $m_Y$ and $m_N$, are presented in Fig.~\ref{fig:m_error}.
For $m_N \gtrsim 50~\GeV$ the measurement deviations of $m_Y$ and $m_N$ can be less than $\sim 2.0~\GeV$.
Generally, the measurement precision for $m_Y$ is better than that for $m_N$.

\begin{figure}[!htbp]
\centering
\subfigure[~Measurement deviation of $m_Y$]{
\includegraphics[width=0.48\textwidth]{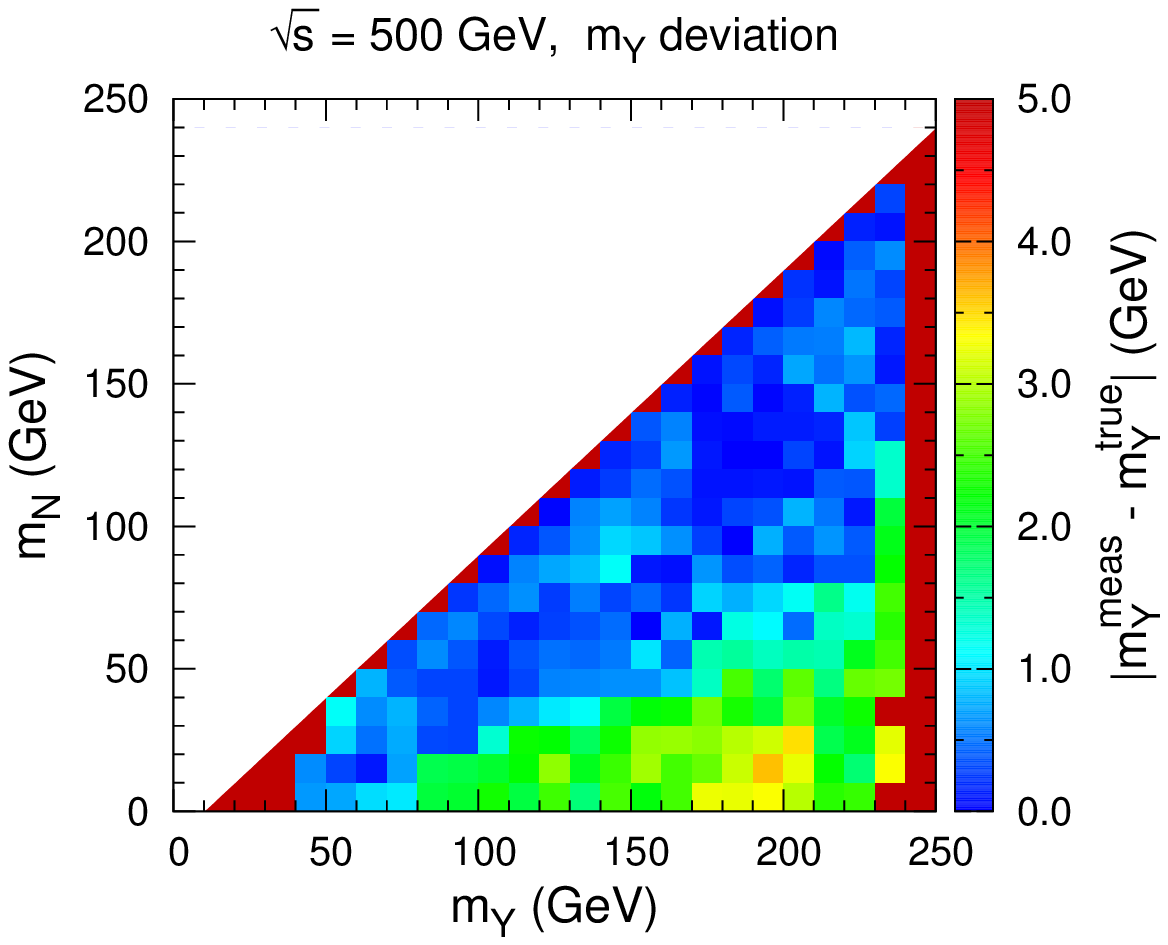}}
\subfigure[~Measurement deviation of $m_N$]{
\includegraphics[width=0.48\textwidth]{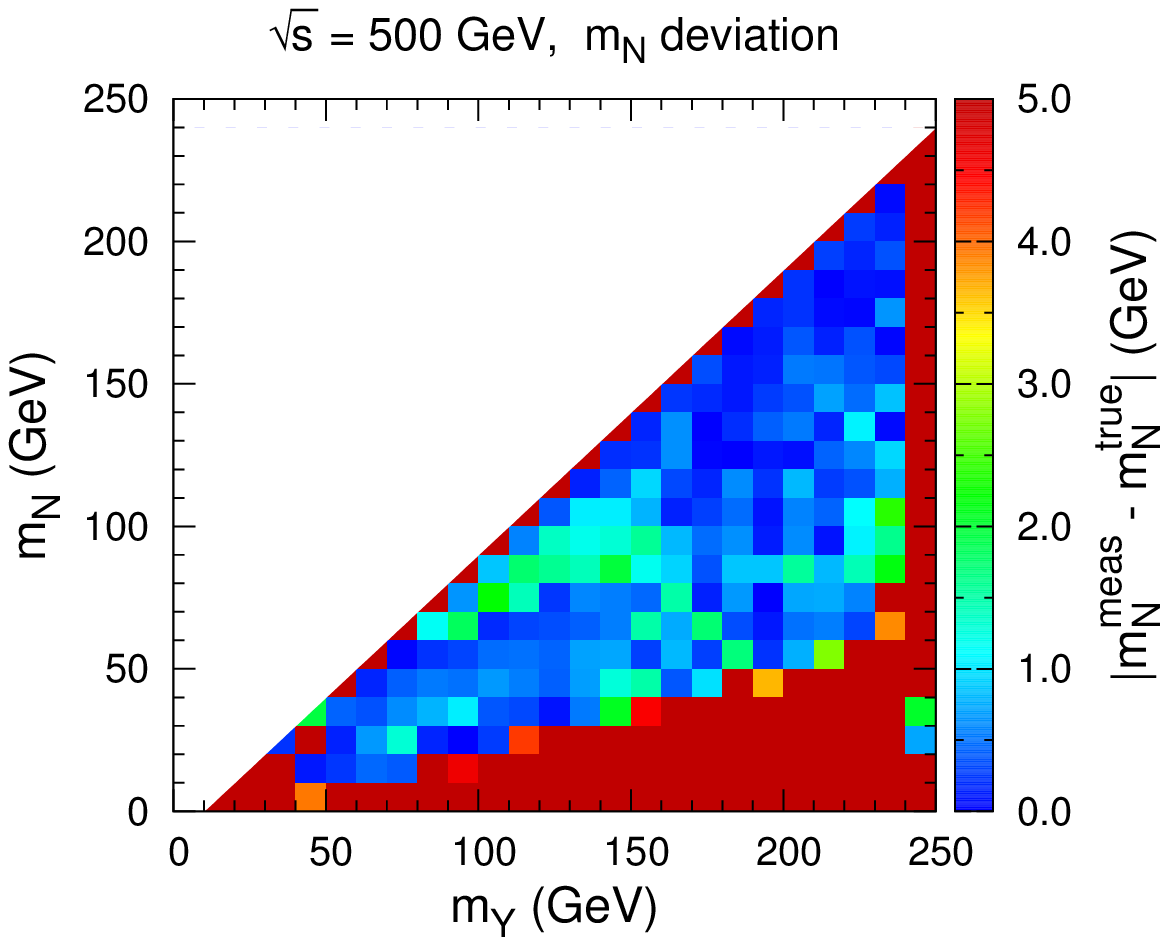}}
\subfigure[~Measurement uncertainty of $m_Y$]{
\includegraphics[width=0.48\textwidth]{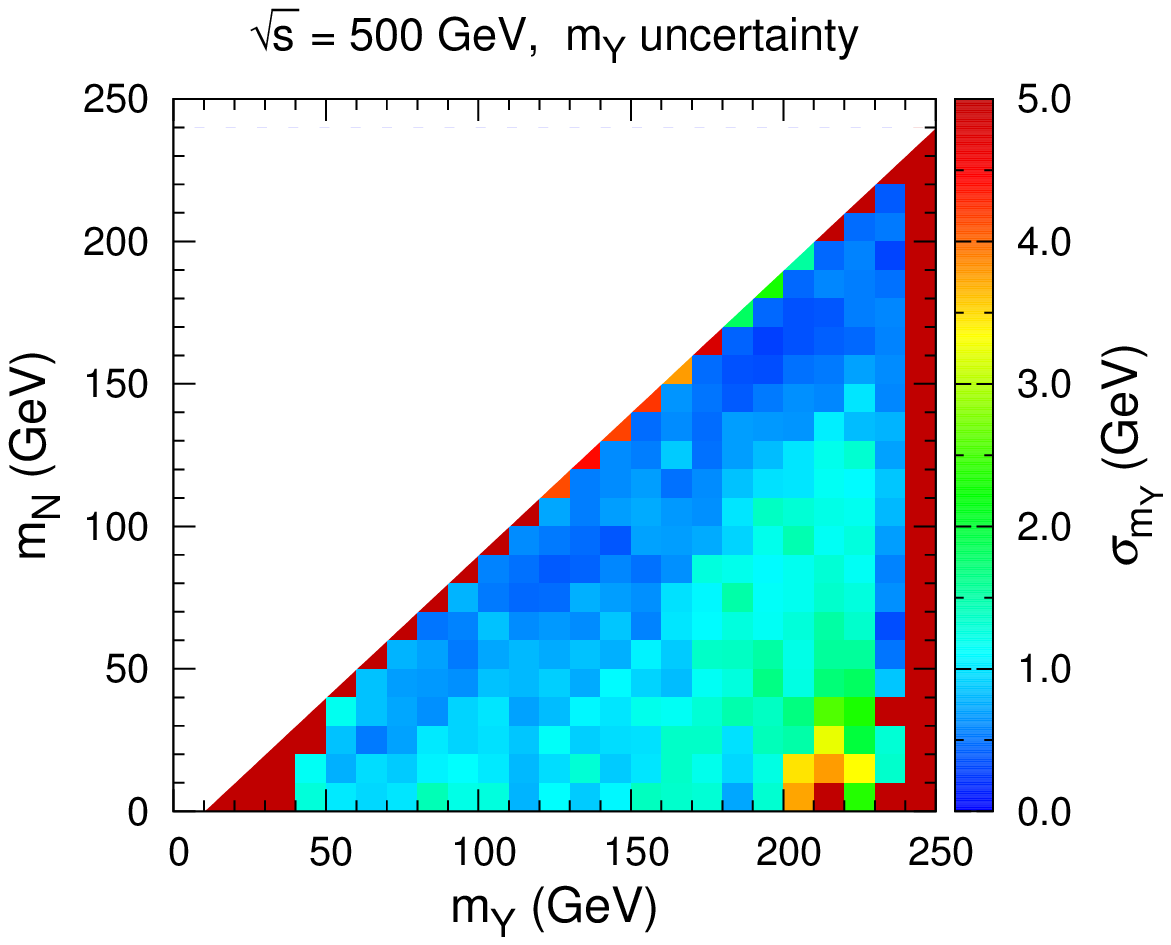}}
\subfigure[~Measurement uncertainty of $m_N$]{
\includegraphics[width=0.48\textwidth]{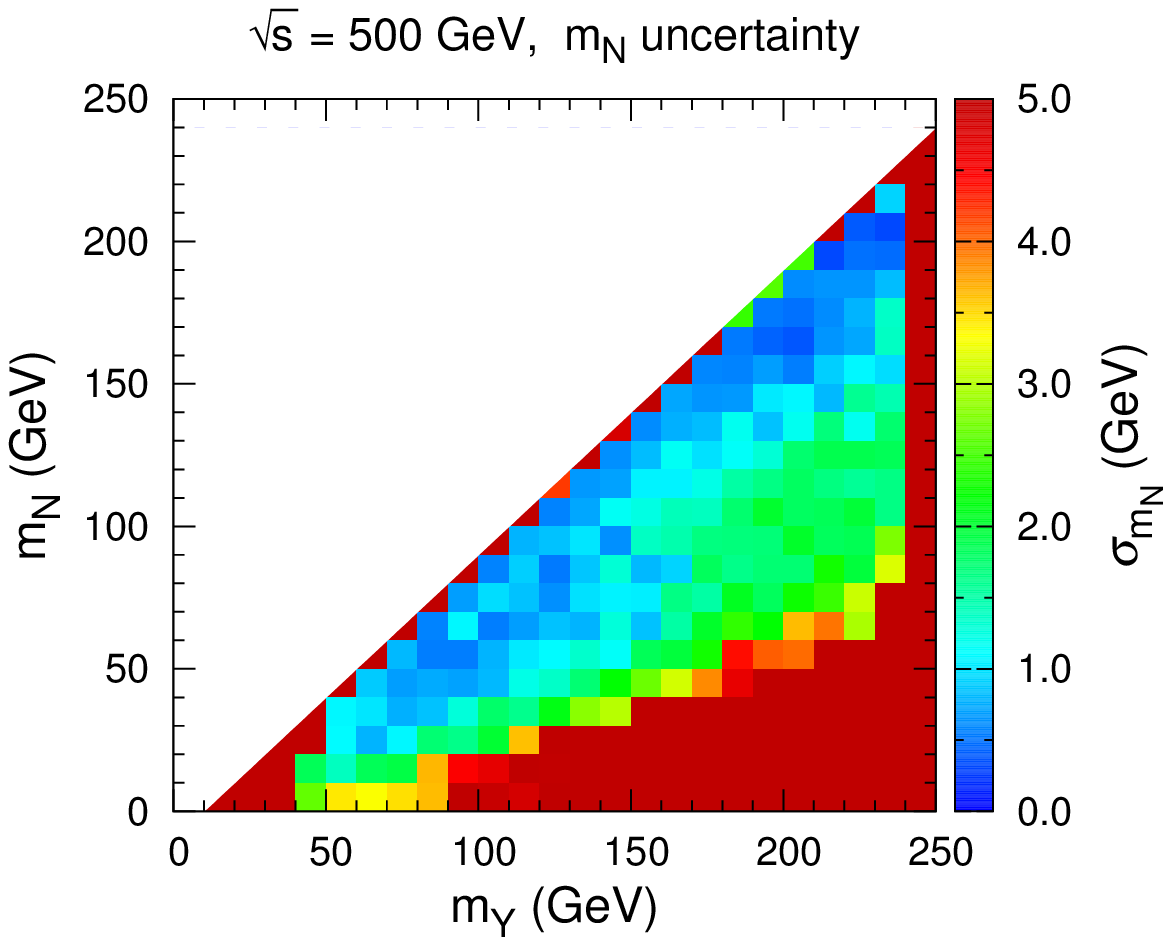}}
\caption{Measurement deviations and uncertainties of $m_Y$ and $m_N$ on the $m_Y$-$m_N$ plane assuming an integrated luminosity of $500~\ifb$ and $\sqrt{s}=500~\GeV$.
The heat map in the top left (top right) panel demonstrates the values of $|m_Y^\mathrm{meas}-m_Y^\mathrm{true}|$ ($|m_N^\mathrm{meas}-m_N^\mathrm{true}|$), while the heat map in the bottom left (bottom right) panel show the uncertainties of the measured $m_Y$ ($m_N$). }
\label{fig:m_error}
\end{figure}

There are three cases where the measurement deviation of $m_N$ could be larger than $5~\GeV$.
For a small $m_N$, such as $m_N \lesssim 40~\GeV$, the solvable region for each event is typically squeezed and the calculated $(\mY, \mN)$ point is probably not in the vicinity of $(m_Y^\mathrm{true}, m_N^\mathrm{true})$.
In addition, for $m_Y \sim 250~\GeV$ the $\smur^+ \smur^-$ production rate is highly suppressed and the requirement of $\mY < 240~\GeV$ is unsuitable.
Moreover, if $m_Y \sim m_N$, the visible particles in the final state are too soft to be reconstructed or well measured.

\begin{figure}[!htbp]
\centering
\subfigure[~Measurement deviation of $m_Y$]{
\includegraphics[width=0.48\textwidth]{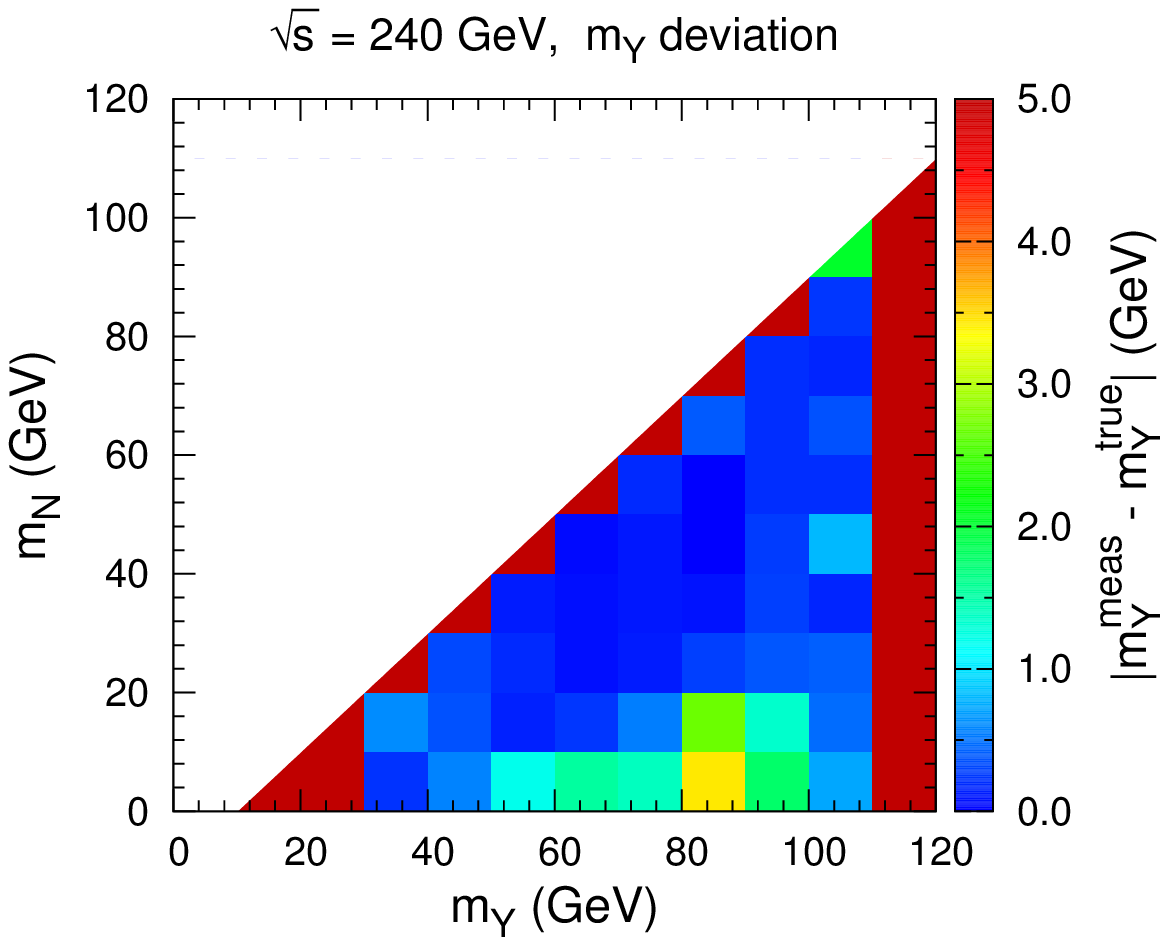}}
\subfigure[~Measurement deviation of $m_N$]{
\includegraphics[width=0.48\textwidth]{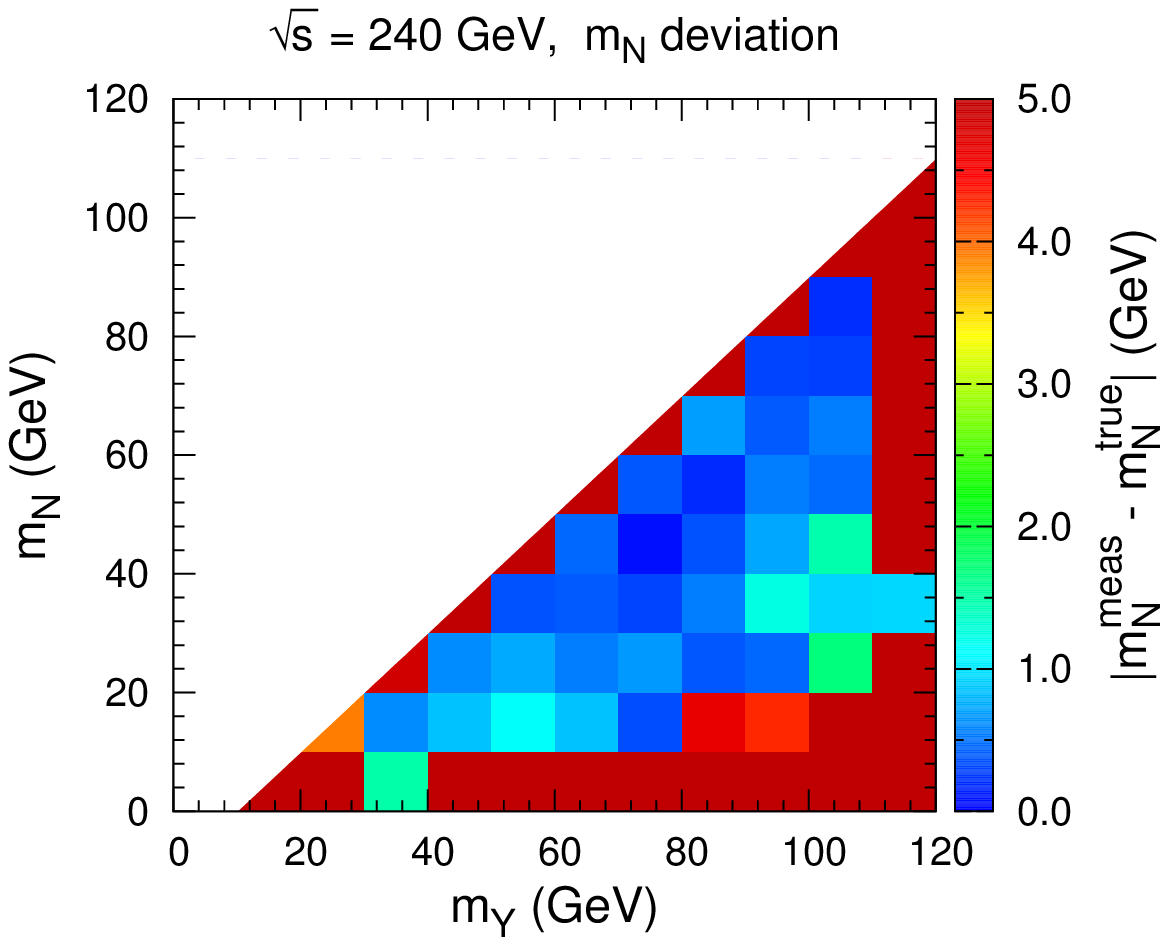}}
\subfigure[~Measurement uncertainty of $m_Y$]{
\includegraphics[width=0.48\textwidth]{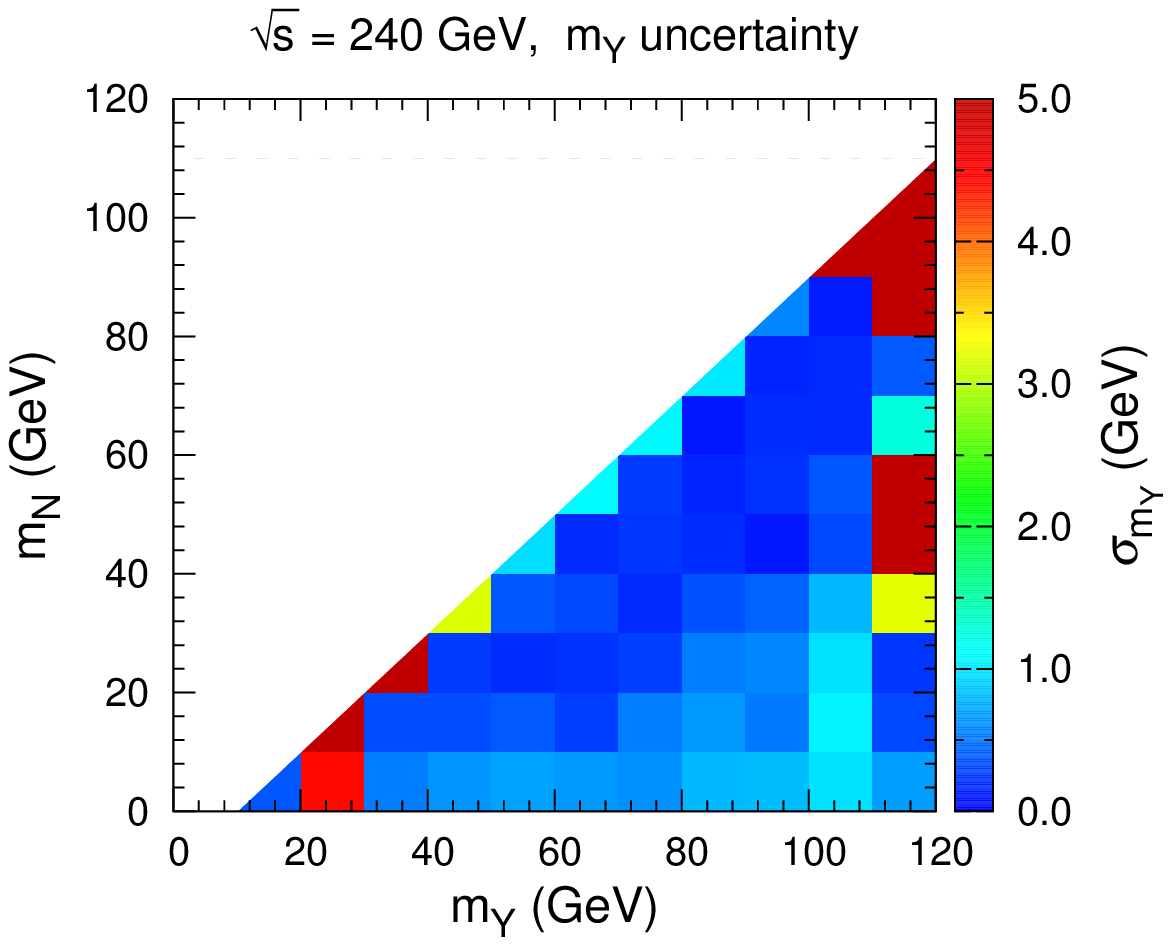}}
\subfigure[~Measurement uncertainty of $m_N$]{
\includegraphics[width=0.48\textwidth]{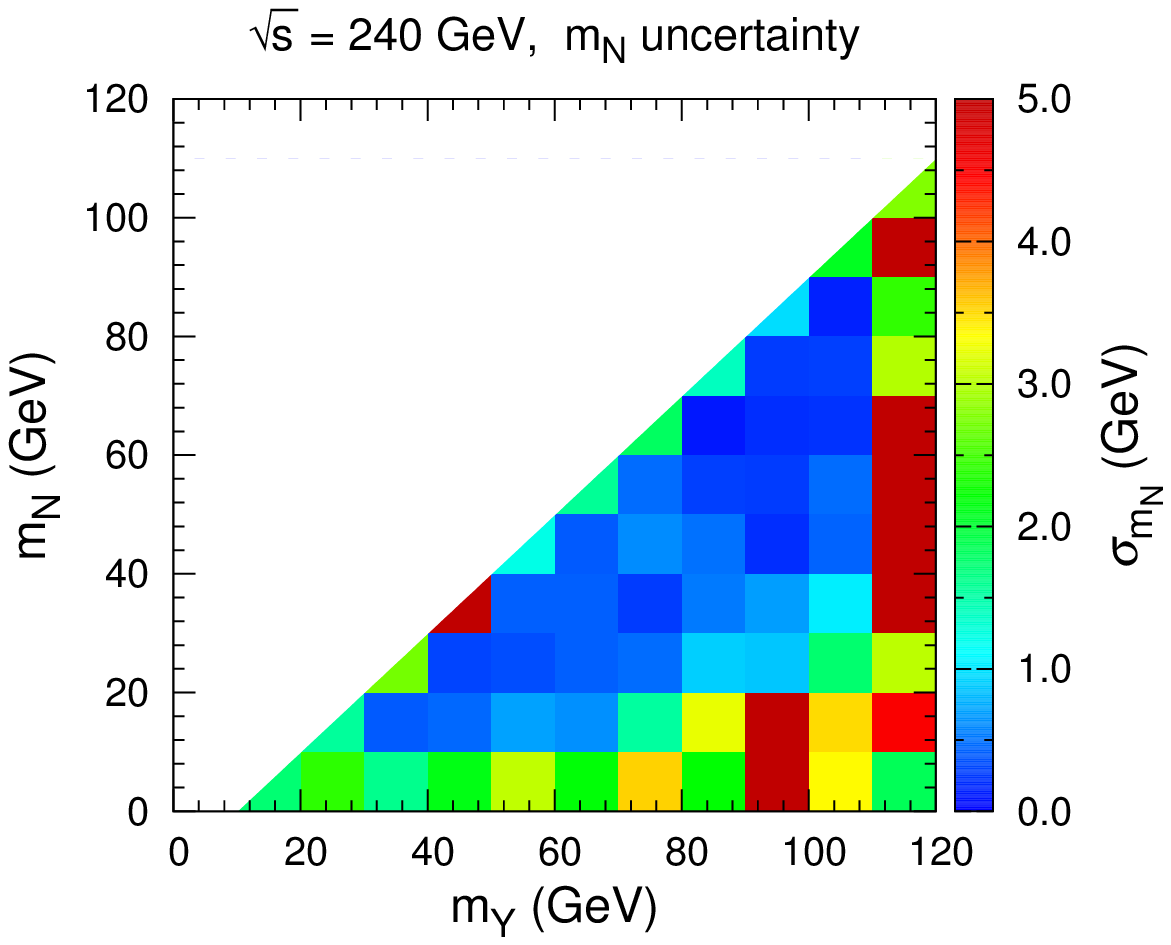}}
\caption{Measurement deviations and uncertainties of $m_Y$ and $m_N$ on the $m_Y$-$m_N$ plane assuming an integrated luminosity of $500~\ifb$ and $\sqrt{s}=240~\GeV$.
The heat map in the top left (top right) panel demonstrates the values of $|m_Y^\mathrm{meas}-m_Y^\mathrm{true}|$ ($|m_N^\mathrm{meas}-m_N^\mathrm{true}|$), while that in the bottom left (bottom right) panel show the uncertainties of $m_Y$ ($m_N$).}
\label{fig:240_m_error}
\end{figure}

Finally, we show the measurement deviations and uncertainties of $m_Y$ and $m_N$ for an $e^+e^-$ collider at $\sqrt{s}=240~\GeV$ with $500~\ifb$ of data in Fig.~\ref{fig:240_m_error}.
We find that for $m_Y \lesssim 105~\GeV$ the measurement deviations of $m_Y$ and $m_N$ can be less than $1.0~\GeV$ and $1.5~\GeV$, respectively.
However, such a parameter region has almost been excluded by the LEP searches.
For $m_Y \sim 120~\GeV$, the measurement deviations of $m_Y$ and $m_N$ would be large.
Furthermore, the uncertainties in this case are smaller than those at a $\sqrt{s} = 500~\GeV$ collider due to better statistical uncertainties arising from larger cross sections.

\section{Conclusions and discussions}
\label{sec:cad}

In this work, we propose a mass measurement method for semi-invisible final states at $e^+e^-$ colliders.
We focus on the pair production of new particles, which subsequently decay into two visible and two invisible particles.
Based on the solvability of kinematic equations, we define a pair of new variables ($m_Y^\mathrm{edge}$, $m_N^\mathrm{edge}$), which are as useful as the ($m_Y^\mathrm{max}$, $m_N^\mathrm{max}$) variables that were proposed in Ref.~\cite{HarlandLang:2012gn}.
It is observed that the event density tends to be the largest near the true mass point on the $\mY$-$\mN$ plane.
By utilizing this fact, we propose a method to extract the physical masses $m_Y$ and $m_N$.

We consider a supersymmetric process $e^+e^-\to \smur^+\smur^- \to \mu^+\mu^-\tchi_1^0\tchi_1^0$ as an illuminating example, although the method only depends on the topology rather than any specified new physics process.
In order to use our mass measurement method in realistic situations,
we take into account detector and background effects in the simulation and calculation.
After imposing some simple cuts, we can efficiently suppress SM backgrounds. We find that background events uniformly scatter on the $\mY$-$\mN$ plane, while signal events would concentrate in the vicinity of the true mass point.

We also introduce an algorithm to further improve the mass measurement by utilizing the shape of the event distribution on the $\mY$-$\mN$ plane.
By comparing the measured masses with the true masses, we show the deviation of the mass measurement for different $m_Y$ and $m_N$.
With an integrated luminosity of $500~\ifb$ and $\sqrt{s}=500$~GeV, we find that the measurement deviation of $m_Y$ ($m_N$) is less than $2.5~(2.0)~\GeV$ for $m_N \gtrsim 50~\GeV$.
These results could be further improved if more data are available and more dedicated cuts are adopted.

Note that some powerful mass measurement methods at $e^+e^-$ colliders have been proposed, and they could have very high precisions. For instance,
the authors of Ref.~\cite{Christensen:2014yya} achieved a precision of $\sim 0.5$ GeV by analyzing the kinematic cusps and endpoints of some kinematic variables. The
threshold scan method by varying the collision energy may be more powerful. The precision can even reach $\sim 0.2$ GeV as shown in Refs.~\cite{Brau:2012hv,Bagger:2000iu}. However, this strategy would not be adopted in early runs of an $e^+e^-$ collider. Here we would like to mention that
the different mass measurement methods can be complementary. The method used in this paper is easy to handle, and
would be very precise for some benchmark points. Moreover, as the new variables are bounded by the true masses,
their distributions can be effectively used to separate the signal events from backgrounds. This issue has been
pointed out in Ref.~\cite{HarlandLang:2012gn}, and can be seen from our Fig.~\ref{fig:dist}.

As mentioned in Ref.~\cite{HarlandLang:2012gn}, the mass measurement methods used at lepton colliders may also be valid at hadron colliders. It is possible to acquire the total 4-momenta of the system for some processes at hadron colliders.
For instance, new particles can be produced in a pair from a central exclusive production process $pp\to pp\gamma\gamma$ with $\gamma\gamma \to Y\bar{Y}$ (see Ref.~\cite{Albrow:2010yb} and references therein).
By installing some proton-tagging detectors far from the interaction point, one would be able to measure the full kinematic information of the two protons in the final state~\cite{Albrow:2008pn,HarlandLang:2011ih}, and then get the full 4-momenta of the $Y \bar{Y}$ system.
In this case, we could extend our method to hadron colliders.

\begin{acknowledgments}
QSY and ZHY acknowledge Yang Bai for early collaboration on related topics.
This work is supported by the National Natural Science Foundation of China under Grants Nos.~11475189, 11475191, 11135009, the 973 Program of China under Grant No.~2013CB837000, and
by the Strategic Priority Research Program
``The Emergence of Cosmological Structures'' of the Chinese
Academy of Sciences under Grant No.~XDB09000000.
ZHY is supported by the Australian Research Council.
\end{acknowledgments}

\appendix

\section{Algorithm for extracting $(m_Y^\mathrm{meas}, m_N^\mathrm{meas})$}
\label{sec:ext}

As discussed in Subsection~\ref{sec:mass_meas}, the true mass point is very likely to lie within the grid with the highest refined density.
An important issue is to extract the measured values of $m_Y$ and $m_N$ with a high precision.
Here we provide an algorithm for it by shrinking the grids.

First, we divide the $m_Y$-$m_N$ plane into grids with a grid width of 3~GeV and calculate the refined density in each grid.
We denote the center of the grid with the highest refined density as $(m_Y^{(1)}, m_N^{(1)})$.
Then we shrink the grid size by increasing both the row and column numbers by one, and again, $(m_Y^{(2)}, m_N^{(2)})$ is defined by the grid with the highest density.
Repeat this procedure and we will obtain a series of $(m_Y^{(j)}, m_N^{(j)})$ with $j=1,2,\cdots$.

As $j$ increases, the grid size becomes smaller and smaller. We should end the algorithm by a criterion judging whether the precision is high enough. For two successive steps $j$ and $j+1$, we examine whether the inequality
\begin{equation}
\sqrt{({m_Y^{(j+1)}-m_Y^{(j)}})^2 + ({m_N^{(j+1)}-m_N^{(j)}})^2} < \sqrt{2} L_j
\label{eq:c}
\end{equation}
is true. Here $L_j$ is the grid width in step $j$.
Fig.~\ref{fig:grid_c} shows two typical cases for the criterion.
If the inequality is true, as demonstrated in Fig.~\ref{fig:grid_c:acc}, the grid with the highest density in step $j+1$ would be likely to contain the true mass point, and we continue to shrink the grids.
If not, as illustrated in Fig.~\ref{fig:grid_c:end}, $(m_Y^{(j)}, m_N^{(j)})$ should be closer to $(m_Y^\mathrm{true}, m_N^\mathrm{true})$, compared with $(m_Y^{(j+1)}, m_N^{(j+1)})$; therefore, we define the measured values $(m_Y^\mathrm{meas}, m_N^\mathrm{meas})=(m_Y^{(j)}, m_N^{(j)})$ and end the algorithm.

\begin{figure}[!htbp]
\centering
\subfigure[~Case A\label{fig:grid_c:acc}]{
\includegraphics[width=0.31\textwidth]{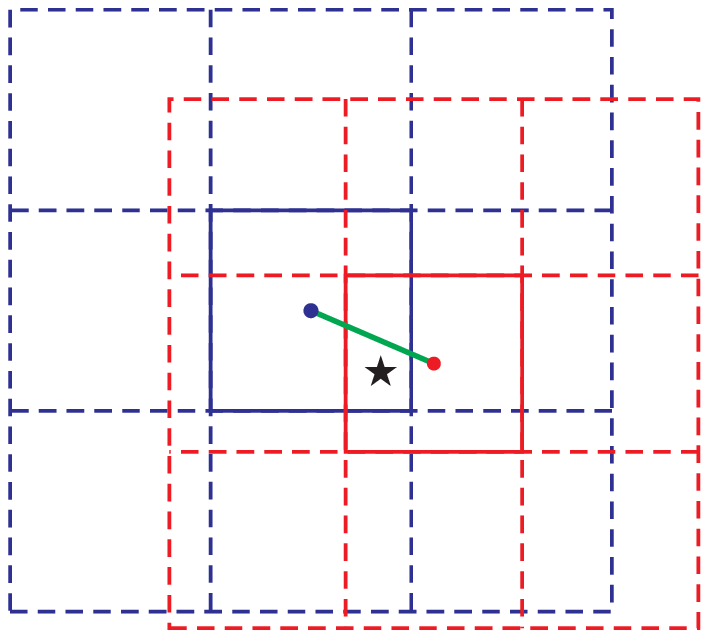}}
\hspace{2em}
\subfigure[~Case B\label{fig:grid_c:end}]{
\includegraphics[width=0.31\textwidth]{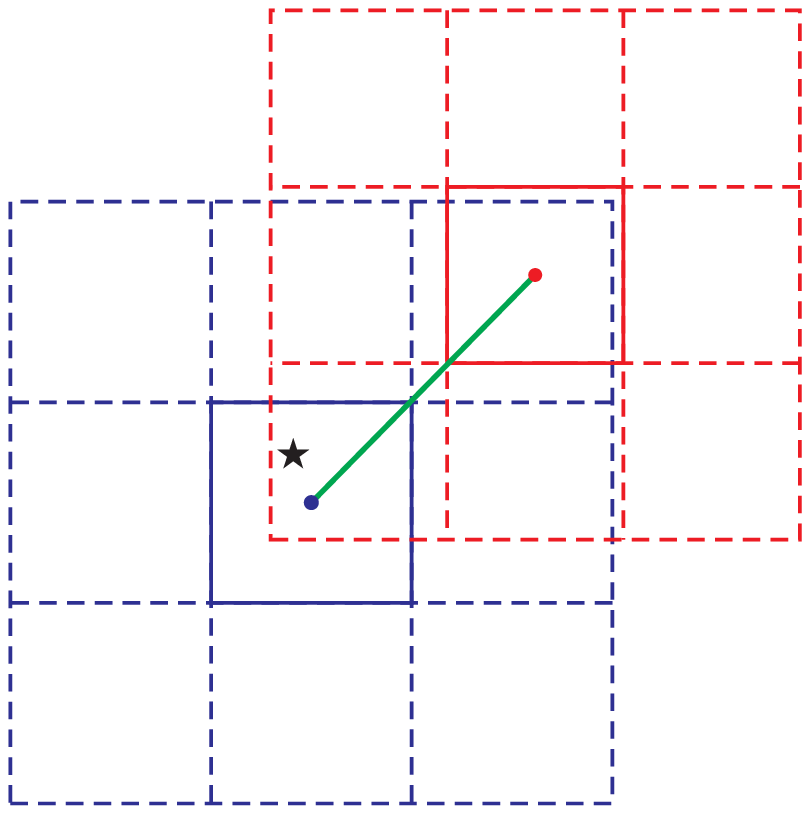}}
\caption{Illustration of two typical cases for the criterion judging if the algorithm should be ended.
The black stars indicate the true mass.
The blue and red grids correspond to step~$j$ and step~$j+1$, respectively.
The solid squares denote the grids with the highest refined density and the blue and red dots correspond to the points $(m_Y^{(j)}, m_N^{(j)})$ and $(m_Y^{(j+1)}, m_N^{(j+1)})$, respectively.
In case A (a), the inequality is satisfied, while in case B (b) the algorithm should be ended.
}
\label{fig:grid_c}
\end{figure}

\bibliographystyle{JHEP}
\bibliography{dmmass}

\end{document}